\documentclass[reprint,prl,aps, floatfix]{revtex4-1}
\usepackage[utf8]{inputenc}
\usepackage[T1]{fontenc}
\usepackage{graphicx}
\usepackage{xcolor}
\usepackage{dcolumn}
\usepackage{bm}
\usepackage{amssymb}
\usepackage{braket}
\usepackage{color,amsmath}
\usepackage{comment}
\usepackage{gensymb}
\usepackage{soul,xcolor} 
\setstcolor{red} 
\usepackage{epstopdf}
\usepackage{hhline}
\usepackage{tabularx}
\usepackage{xspace}
\usepackage{layouts}
\usepackage{lineno} 
\usepackage{xr} 
\usepackage{amsfonts}
\usepackage{bbold}
\usepackage{placeins}

\newcolumntype{C}[1]{>{\centering\arraybackslash}m{#1}}
\newcolumntype{N}{@{}m{0pt}@{}}

\begin{document}

\title{Pervasive spin-triplet superconductivity in rhombohedral graphene}

\author{Manish Kumar$^{1*}$}
\author{Derek Waleffe$^{1*}$} 
\author{Anna Okounkova$^{1*}$}
\author{Raveel Tejani$^{2,3}$}
\author{Kenji Watanabe$^{6}$}
\author{Takashi Taniguchi$^{7}$}
\author{\'Etienne Lantagne-Hurtubise$^{8}$}
\author{Joshua Folk$^{2,3\dagger}$}
\author{Matthew Yankowitz$^{1,9\dagger}$}

\affiliation{$^{1}$Department of Physics, University of Washington, Seattle, Washington, 98195, USA}
\affiliation{$^{2}$Department of Physics and Astronomy, University of British Columbia, Vancouver, British Columbia, V6T 1Z1, Canada}
\affiliation{$^{3}$Quantum Matter Institute, University of British Columbia, Vancouver, British Columbia, V6T 1Z1, Canada}
\affiliation{$^{6}$Research Center for Electronic and Optical Materials, National Institute for Materials Science, 1-1 Namiki, Tsukuba 305-0044, Japan}
\affiliation{$^{7}$Research Center for Materials Nanoarchitectonics, National Institute for Materials Science, 1-1 Namiki, Tsukuba 305-0044, Japan}
\affiliation{$^{8}$Département de physique, Institut quantique $\&$ RQMP, Université de Sherbrooke, Sherbrooke, Québec J1K 2R1, Canada}
\affiliation{$^{9}$Department of Materials Science and Engineering, University of Washington, Seattle, Washington, 98195, USA}
\affiliation{$^{*}$These authors contributed equally to this work.}
\affiliation{$^{\dagger}$jfolk@physics.ubc.ca (J.F.) and myank@uw.edu (M.Y.)}

\maketitle

\textbf{Magnetic fields typically suppress superconductivity once the Zeeman energy exceeds the pairing gap, unless mechanisms such as unconventional pairing, strong spin–orbit coupling, or intrinsic magnetism intervene. Several graphene platforms realize such mitigating routes, exhibiting superconductivity resilient to magnetic fields. Here we report superconductivity in rhombohedral heptalayer graphene that is both induced and stabilized by in-plane magnetic field ($B_{\parallel}$), with critical fields far beyond the Pauli paramagnetic limit. The superconductivity spans a wide gate range and emerges from a sharp zero-field resistive ridge that tracks approximately constant conduction band filling. The presence of zero-field superconductivity and the evolution of the critical temperature with $B_{\parallel}$ are highly gate sensitive. We also observe a weak superconducting diode effect in several distinct regimes within the superconducting phase, including nearby to an integer quantum anomalous Hall state generated by a boron nitride moir\'e superlattice, indicating a potential coexistence of valley imbalance and superconductivity. These results establish several intriguing new properties of spin-triplet, field-induced superconductivity in a thick rhombohedral graphene stack.}

Superconductivity and magnetism are traditionally antagonists, yet in some quantum materials magnetic fields can protect or even create superconductivity. Several microscopic routes can mitigate Zeeman depairing and allow superconductivity to persist to fields exceeding the Pauli paramagnetic limit~\cite{Clogston1962Pauli,Chandrasekhar1962Pauli}. Examples include spin-orbit coupling that suppresses paramagnetic pair breaking in non-centrosymmetric settings (including ``Ising'' protection in two-dimensional materials)~\cite{Lu2015Ising,Xi2016Ising,Saito2016SpinValley,Smidman2017}, finite-momentum pairing that accommodates Fermi-surface mismatch when orbital depairing is weak (Fulde--Ferrell--Larkin--Ovchinnikov mechanism)~\cite{FuldeFerrell1964,LarkinOvchinnikov1964,MatsudaShimahara2007}, and exchange fields from ordered moments that compensate spin splitting (Jaccarino--Peter effect)~\cite{JaccarinoPeter1962,Meul1984}. Apparent exceedance of the weak-coupling Pauli limit can also arise from a reduced or anisotropic quasiparticle $g$-factor~\cite{Clogston1962Pauli,Chandrasekhar1962Pauli,Altarawneh2012UniaxialAnisotropy}, strong-coupling~\cite{Carbotte1990BosonExchange,Orlando1979CriticalFields} or multiband effects~\cite{Gurevich2003TwoGapHc2}, and spin-orbit scattering in the dirty limit~\cite{Werthamer1966WHH,Klemm1975KLB}. In quasi-two-dimensional materials, orbital effects of in-plane magnetic fields are suppressed, so the fate of superconductivity becomes sensitive to the balance among Zeeman and spin-orbit couplings (SOC), band dispersion, and pairing interactions.

\begin{figure*}[t]
\includegraphics[width=\textwidth]{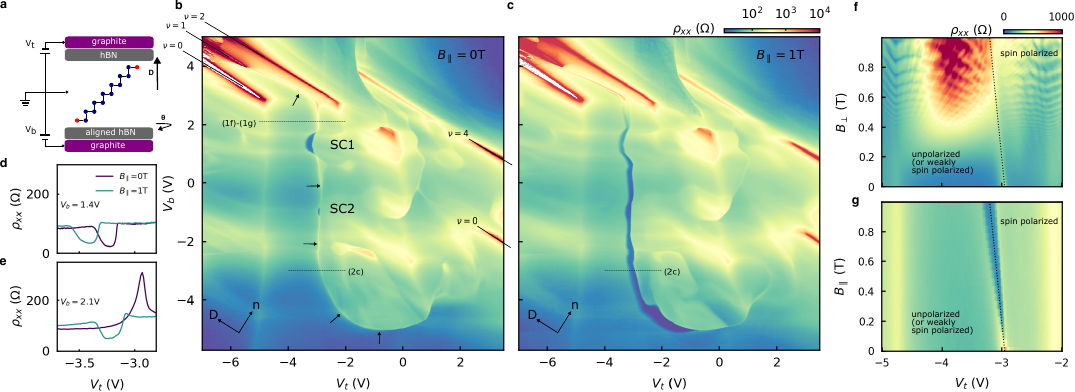} 
\caption{\textbf{Spin-triplet superconductivity in moir\'e R7G induced by $B_{\parallel}$.}
\textbf{a}, Schematic of the rhombohedral heptalayer graphene device with a $13.5$~nm moir\'e superlattice from alignment with the bottom hBN.
\textbf{b}, Measurements of $\rho_{xx}$ versus $V_t$ at fixed $V_b=1.4$~V with $B_{\parallel}=0$~and $1$~T.
\textbf{c}, Same with $V_b=2.1$~V
\textbf{d}, Map of $\rho_{xx}$ versus $V_b$ and $V_t$ at zero field. Arrows denote the sharp resistive feature discussed in the text. Diagonal lines denote selected integer band filling factors, $\nu$.
\textbf{e}, Same map at $B_{\parallel}=1$~T.
\textbf{f}, Landau fan of $\rho_{xx}$ taken versus $V_t$ and $B_{\perp}$ with $B_{\parallel}=0$ and fixed $V_b=2.1$~V.
\textbf{g}, Same measurement taken versus $B_{\parallel}$ at $B_{\perp}=0$. The black dashed lines in \textbf{f} and \textbf{g} are identical guides for the eye.
}
\label{fig:1}
\end{figure*}

Graphene platforms provide a clean venue to dissect these ingredients because the polarization of spin, valley, and layer degrees of freedom are gate tunable. In Bernal bilayer graphene, superconductivity can be induced by a small in-plane field near isospin-ordered metallic phases~\cite{Zhou2022_BBG}. Rhombohedral trilayer graphene hosts two superconducting regimes, one of which is consistent with spin-triplet yet valley-unpolarized pairing and exhibits critical fields well beyond the weak-coupling Pauli limit~\cite{Zhou2021_RTGSC}. Mirror-symmetric twisted trilayer graphene likewise shows pronounced resilience to $B_{\parallel}$~\cite{Cao2021_MATTG}. In thicker rhombohedral multilayers, signatures of chiral superconductivity have been reported, suggestive of odd-parity pairing with both spin and valley polarization~\cite{Han2024, Morisette2025SC}. Across these systems, quasi-2D flat bands and interaction-driven isospin ordering combine to stabilize unconventional superconductivity~\cite{Cao2018, Yankowitz2019, Lu2019_Efetov, Cao2021_MATTG, Hao2021_Phillip, Park2022_Pablo, Zhang2022_NadjPerge, Lin2022_Leo, Zhang2023, Li2024, Holleis2025, Patterson2025, Choi2025, Zhang2025, Yang2025, Morisette2025SC, Han2024, Yang2025_MagSC, Seo2025_UncSC, Kumar2025_dual, Deng2025_Xiaomeng} including spin-triplet or finite-momentum pairing.

Here, we extend these studies to rhombohedral heptalayer graphene (R7G) aligned on one side to hexagonal boron nitride (Fig.~\ref{fig:1}a), which hosts two superconducting pockets and an integer quantum anomalous Hall state at zero magnetic field~\cite{Kumar2025_dual}. These features lie near a sharp resistive ridge that drifts through the $(V_b,V_t)$ plane, where $V_b$ and $V_t$ are the bottom and top gate voltages. We show that a modest in-plane field transforms extended segments of this ridge into superconducting regions. In essence this behavior echoes Bernal bilayer phenomenology, but the surface-bifurcated low-energy states in thicker rhombohedral graphene (analogous to the Su–Schrieffer–Heeger (SSH) chain~\cite{Su1979,Xiao2011,Guinea2006,Shi2020}) introduce an additional surface degree of freedom that reshapes electronic screening. We discuss mechanisms consistent with the phase diagram, including considerations of intrinsic spin–orbit coupling, a Hund's spin exchange term, Zeeman energy from an in-plane magnetic field ($B_{\parallel}$), and a general tendency toward isospin polarization when the density of states is large.

\medskip\noindent\textbf{Superconductivity induced by $B_{\parallel}$}

We first present measurements of the longitudinal resistivity ($\rho_{xx}$) versus $V_b$ and $V_t$ at $B=0$ (Fig.~\ref{fig:1}d) and $B_{\parallel}=1$~T (Fig.~\ref{fig:1}e). At $B=0$, a sharp resistive peak (marked by arrows in Fig.~\ref{fig:1}d) moves through the gate map, hosting two small pockets of superconductivity, SC1 and SC2. When $B_\parallel$ is raised to $1$~T, however, the deep blue color signifying low resistance extends along the entire length of the feature. This feature   
follows approximately constant conduction-band filling, an interpretation explained in detail in Ref.~\cite{Kumar2025_dual} and inferred from its near-perfect tracking of lines of constant carrier density when conduction and valence band are separated ($V_b>2.8$~V), then evolving to track vertically when the bands overlap.

Figure~\ref{fig:1}b shows a linecut from these colormaps at $V_b=1.4$~V, representing the evolution of $\rho_{xx}$ versus $V_t$ in the center of the SC1 pocket as $B_{\parallel}$ is raised from 0 to $1$~T. The sharp dip in $\rho_{xx}$ that reflects superconductivity drifts to more negative $V_t$ as $B_{\parallel}$ increases. Figure~\ref{fig:1}c shows the analogous measurement at $V_b=2.1$~V. In this case, a sharp resistive peak at $B=0$ becomes a large dip at $B_{\parallel}=1$~T, located slightly to its left. As shown below, this dip corresponds to field-induced, spin-triplet superconductivity emerging from the sharp resistive peak.

Low-energy electronic states in graphene carry fourfold degeneracy arising from spin and valley. The valley pseudospin is strongly anisotropic (Ising-like), with a quantization axis normal to the plane, whereas spin is isotropic and can rotate freely to align with an external magnetic field in the absence of SOC. To assess the isospin polarization of the metallic states neighboring superconductivity, we measure $\rho_{xx}$ as functions of both $B_{\perp}$ and $B_{\parallel}$ while sweeping $V_t$ across the sharp resistive peak at fixed $V_b=2.1$~V (Figs.~\ref{fig:1}f,g). As $B_{\parallel}$ increases, the peak shifts and evolves into the sharp dip that signals superconductivity. The nearly identical slopes of these features with $B_{\perp}$ and $B_{\parallel}$ (for which orbital couplings differ greatly) indicate that Zeeman energy is the relevant scale. This is consistent with a (fully or partially) spin-polarized phase to the right gaining Zeeman energy relative to a weakly polarized (or unpolarized) phase to the left (quantum-oscillation analysis supports this interpretation; Extended Data Fig.~\ref{fig:PRB_FFT}).

\begin{figure*}[t]
\includegraphics[width=0.8\textwidth]{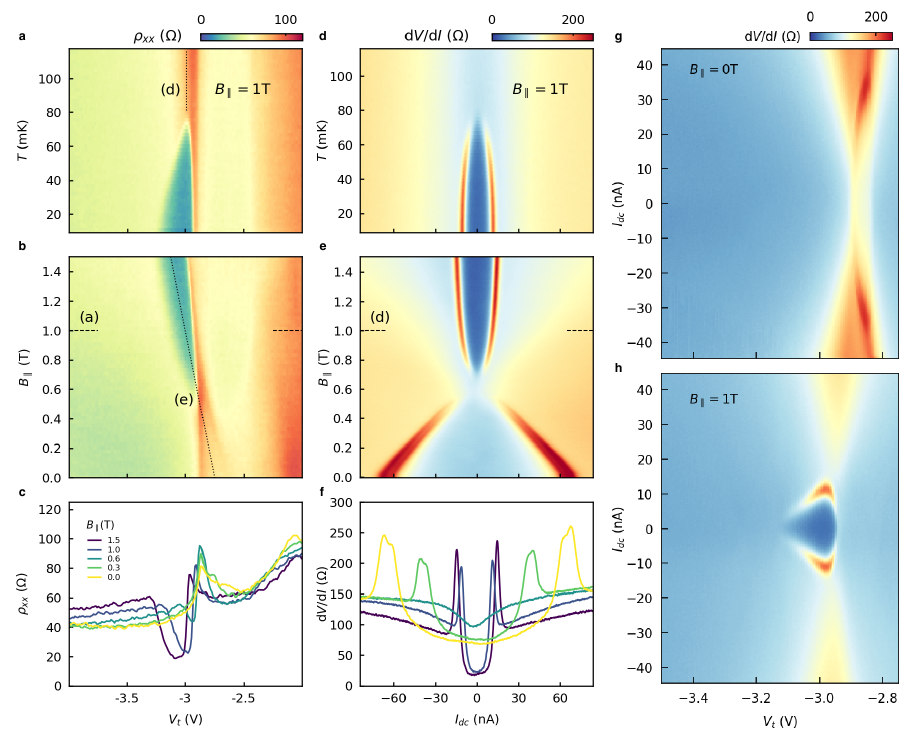} 
\caption{\textbf{Nonlinearities associated with states along the sharp resistive bump.}
\textbf{a}, $\rho_{xx}$ versus $V_t$ and $T$ at $B_\parallel=1$~T and $V_b=-3$~V.
\textbf{b}, $\rho_{xx}$ versus $V_t$ and $B_{\parallel}$ at $V_b=-3$~V.
\textbf{c}, Selected traces from \textbf{(b)}.
\textbf{d}, d$V$/d$I$ versus $I_{dc}$ and $T$ at $B_\parallel=1$~T, $V_b=-3$~V and $V_t=-3$~V (black dashed line in \textbf{(a)}).
\textbf{e}, d$V$/d$I$ versus $I_{dc}$ and $B_{\parallel}$ taken along the trajectory of the tilted black dashed line in \textbf{b}.
\textbf{f}, Selected traces from \textbf{(e)}. 
\textbf{g}, d$V$/d$I$ versus $V_t$ and $I_{dc}$ taken at $V_b=0$~V and $B_{\parallel}=0$~T. 
\textbf{h}, Same as \textbf{(h)} with $B_{\parallel}=1$~T.
}
\label{fig:2}
\end{figure*}

We analyze the emergence of superconductivity with $B_{\parallel}$ in greater detail using the measurements in Fig.~\ref{fig:2}a-c, taken along the black dashed trajectory in Fig.~\ref{fig:1}d ($V_b=-3.0$~V). At $B_\parallel=1$~T, the temperature dependence of resistivity exhibits a characteristic superconducting dome (Fig.~\ref{fig:2}a). Maps of $\rho_{xx}$ versus $B_\parallel$ (Figs.~\ref{fig:2}b,c) show that the resistive bump does not shift with field up to 0.5~T, then for $B_\parallel>0.6$~T it shifts to the left as the sharp dip associated with superconductivity emerges (similar behavior recurs across many gate voltages; Extended Data Fig.~\ref{fig:BparVtMaps}). Figures~\ref{fig:2}e,f show the differential resistivity (d$V$/d$I$) versus dc current ($I_{dc}$) and $B_{\parallel}$ along the diagonal trajectory marked by the dashed black line in Fig.~\ref{fig:2}b. In addition to the characteristic nonlinearities associated with superconductivity at large $B_{\parallel}$ (Fig.~\ref{fig:2}d), the non-superconducting state at lower field also exhibits pronounced nonlinearities. At $B=0$ there are sharp d$V$/d$I$ peaks at $I_{dc}\approx\pm65$~nA surrounding a lower resistance state at low bias. As $B_{\parallel}$ is raised, the d$V$/d$I$ peaks move towards zero and disappear near $B_{\parallel}\approx0.6$~T (similar behavior occurs with $B_{\perp}$, see Extended Data Fig.~\ref{fig:Non-linearity}a,b). The d$V$/d$I$ peaks reappear at larger $B_{\parallel}$, flanking a deep suppression of d$V$/d$I$ at small $I_{dc}$ when the dashed black trajectory in Fig.~\ref{fig:1}b crosses over into the superconducting state. Crucially, d$V$/d$I$ nonlinearities occur only for $V_t$ values that host the elevated-resistance feature at small $B_\parallel$ and superconductivity at larger field. Transport at small $I_{dc}$ remains linear in the surrounding metallic phases (Figs.~\ref{fig:2}g,h, taken at $V_b=0$), irrespective of their different isospin polarizations. 

\begin{figure*}[t]
\includegraphics[width=\textwidth]{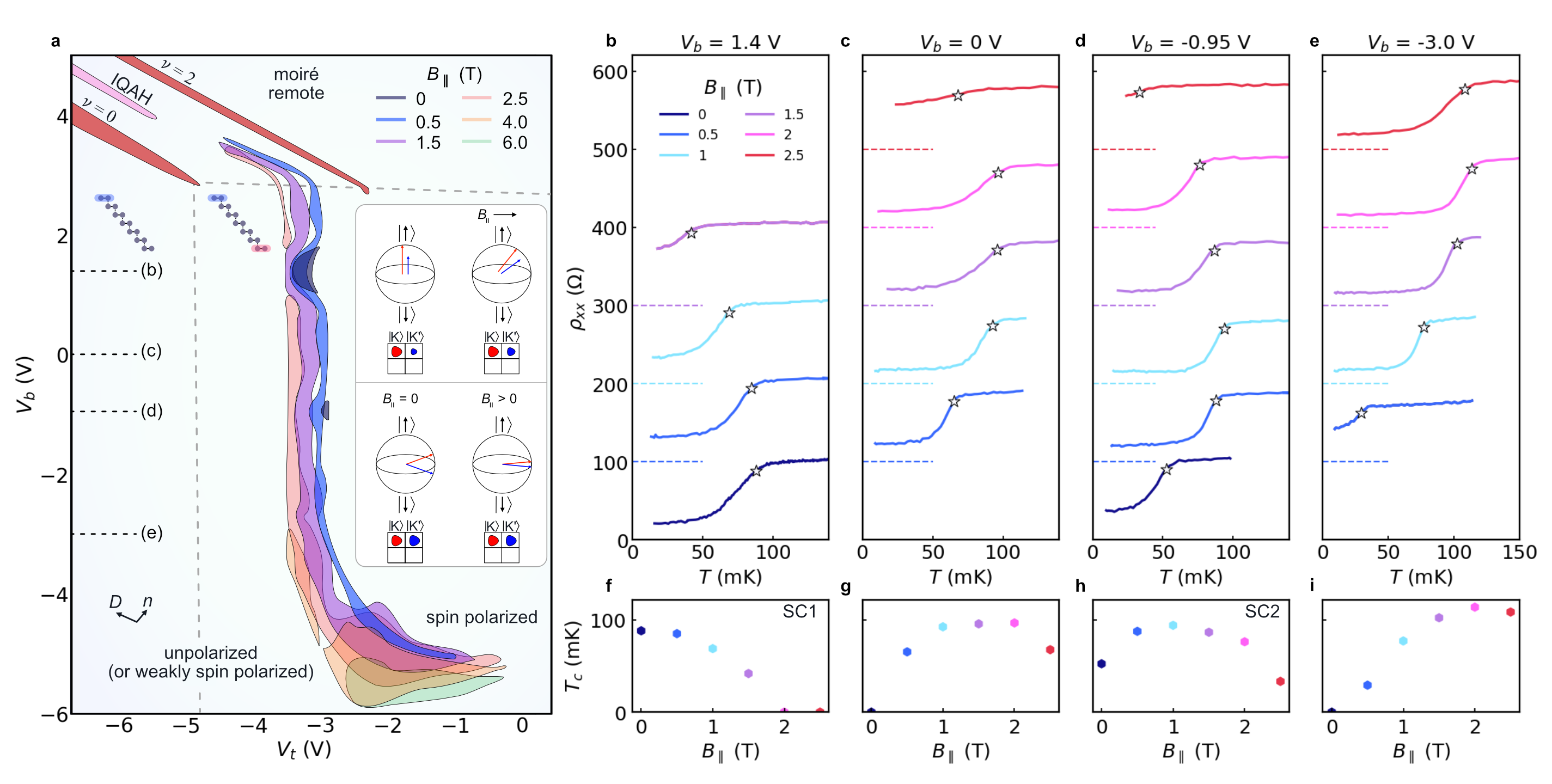} 
\caption{\textbf{Non-monotonic evolution of $T_c$ with $B_{\parallel}$.}
\textbf{a}, Schematic phase diagram denoting the regions of the $(V_b,V_t)$ plane exhibiting superconductivity at various selected $B_{\parallel}$. Red (pink) regions denote trivial (Chern) insulating states. (inset) Bloch spheres show two possible spin-polarized states and their evolution with $B_\parallel$; the red and blue arrows respectively represent the spin polarization in valleys $K$ and $K'$. Fermi surfaces associated with the different valleys are depicted in red and blue in the Punnett square diagrams. The top half of the inset depicts an easy-axis state with weak valley imbalance, and the bottom half shows a valley-balanced spin canted state.
\textbf{b}, $\rho_{xx}$ versus $T$ measurements at several selected values of $B_{\parallel}$ with $B_{\perp}=0$. The measurement is taken with $V_b=1.4$~V and optimal $V_t$. Curves are vertically offset for clarity. Dashed lines denote $\rho_{xx}=0$ for each curve of corresponding color. Stars denote $T_c$.
\textbf{c}, Same for $V_b=0$ V. 
\textbf{d}, Same for $V_b=-0.95$~V.
\textbf{e}, Same for $V_b=-3.0$~V
\textbf{f-i}, $T_c$ versus $B_{\parallel}$ for each of the four associated values of $V_b$.
}
\label{fig:3}
\end{figure*}

Owing to the narrowness of the resistive bump, we are not able to resolve associated quantum oscillations that would further diagnose its isospin order. Even so, the sharp d$V$/d$I$ peaks constrain feasible origins. One possibility is a charge density wave (CDW) that depins above a critical $I_{dc}$, although CDW sliding typically reduces d$V$/d$I$ rather than producing sharp peaks~\cite{Gruner1988Dynamics}. Another possibility is a state that is weakly valley-polarized (akin to a partially polarized quarter metal or three-quarters metal), in which a dc current exerts a torque that can move orbital-magnetization domain walls~\cite{Tschirhart2023SpinHallTorque}, yielding a sharp $\mathrm{d}V/\mathrm{d}I$ feature. Although we do not observe clear anomalous Hall effect associated with the sharp resistive bump---as typically accompanies valley imbalance---we later show a superconducting diode effect under $B_{\parallel}$ that keeps this scenario in play. Related behavior appears in Bernal bilayer graphene (BBG), which also shows a sharp resistive feature with nonlinear transport at $B=0$ and superconductivity induced from that phase by a small in-plane magnetic field~\cite{Zhou2022_BBG}. It is not yet clear whether the low- and high-field states in our sample and in BBG are identical, but their close correspondence motivates the possibility of a shared origin.

\medskip\noindent\textbf{Non-monotonic evolution of $T_c$ with $B_{\parallel}$}

As shown in Figs.~\ref{fig:1}d, e, two zero-field superconducting pockets (SC1 and SC2~\cite{Kumar2025_dual}) merge into a single elongated strip as $B_{\parallel}$ increases, forming a ``river'' in the $(V_b,V_t)$ plane. The evolution along this river appears to be governed by the layer-dependent charge distribution, an important degree of freedom in rhombohedral multilayers. For small external layer potentials (set by the voltage difference between the two gates), carriers from different bands polarize to opposite crystal surfaces, yielding a strongly interacting semimetal~\cite{Kumar2025_dual}. By contrast, for large layer potentials carriers are confined to either the valence or conduction band and localize exponentially on one surface. In our device, the superconductivity is modulated by this spatial charge configuration.

In the single-surface regime ($V_b>2.8$~V), insulating states and the sharp resistive ridge associated with superconductivity follow diagonal trajectories because both gates tune the conduction band. For $V_b<2.8$~V, where charge resides on both surfaces, gate-screening effects cause features to lock to trajectories controlled primarily by a single gate~\cite{Kolar2025}. Figure~\ref{fig:3}a marks the key features---insulating states and superconducting regions---at selected $B_{\parallel}$ (see also Extended Data Fig.~\ref{fig:FiniteFieldMaps}). The field-induced superconductivity spans distinct charge configurations. For $V_b>2.8$~V it corresponds to conduction band carriers on a single surface. As $V_b$ is reduced, valence band carriers populate the opposite surface and screen the bottom gate. For $V_b<-4.2$~V the superconductivity bends away from fixed $V_t$, possibly indicating a reversal of the internal electric field and a switch of the surface hosting conduction-band states.

The evolution of $T_c$ with $B_{\parallel}$ is complicated and depends sensitively on position along the river. In SC1, which is already present at $B=0$, $T_c$ decreases monotonically with $B_{\parallel}$ (Fig.~\ref{fig:3}b). The behavior differs markedly at gate voltages that instead host a resistive bump at $B=0$. In such cases, $T_c$ grows slowly with $B_{\parallel}$ once superconductivity first appears with field, peaks at several tesla, and then vanishes at higher fields (Figs.~\ref{fig:3}c,e; Extended Data Fig.~\ref{fig:TC_Vb_B}). SC2 shows a weaker version of this trend (Fig.~\ref{fig:3}d). The gradual rise of $T_c$ outside of SC1 is inconsistent with a simple picture in which superconductivity becomes strongest immediately after a competing order is suppressed; instead it points to a more delicate balance among the mechanisms responsible for determining its strength.

\begin{figure*}[t]
\includegraphics[width=\textwidth]{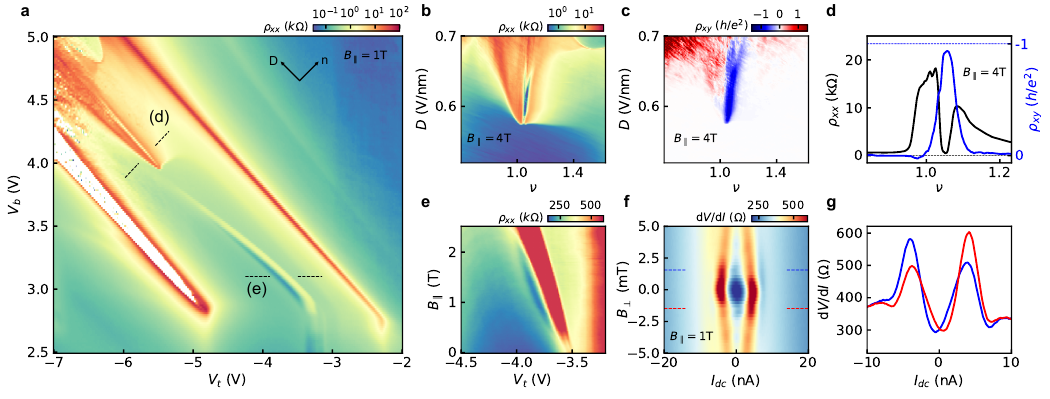} 
\caption{\textbf{Diodic field-induced superconductivity nearby an integer quantum anomalous Hall state.}
\textbf{a}, Zoomed-in map of $\rho_{xx}$ versus $V_b$ and $V_t$ at $B_{\parallel}=1.0$~T. 
\textbf{b}, Zoomed-in map of the IQAH state near $\nu=1$, showing $\rho_{xx}$ versus $\nu$ and $D$ at $B_{\parallel}=4.0$~T.
\textbf{c}, Same as \textbf{(b)}, showing $\rho_{xy}$.
\textbf{d}, Linecuts taken from the data shown in \textbf{(b)} and \textbf{(c)}, taken at the position shown by the tilted dashed line in \textbf{(a)}.
\textbf{e}, $\rho_{xx}$ versus $V_t$ and $B_{\parallel}$ taken at fixed $V_b=3.1$~V (along the trajectory denoted in \textbf{(a)}). 
\textbf{f}, d$V$/d$I$ versus $I_{dc}$ and $B_{\perp}$ taken at $V_b=3.0$~V, $V_t=-3.6$~V and $B_{\parallel}=1.0$~T.
\textbf{g}, Linecuts from \textbf{(f)} taken at $B_\perp\approx\pm1.5$~mT. $\rho_{xx}$ data shown in \textbf{(b)} and \textbf{(d)} are taken using a second contact pair ($V_{xx2}$; see Methods).
}
\label{fig:4}
\end{figure*}

\medskip\noindent\textbf{Superconducting diode effect nearby an IQAH state}

We next analyze transport when a sufficiently large displacement field $D$ polarizes conduction band charges to a single crystal surface, with no coexisting valence-band carriers (i.e., $V_b>2.8$~V). Figure~\ref{fig:4}a shows a zoomed-in map of this region at $B_{\parallel}=1$~T. At the bottom of the map the superconductivity lies around $V_t=-3.3$~V and has essentially no dependence on $V_b$, but turns to move diagonally together with nearby insulators along lines of nearly constant carrier density once the valence band is emptied. Above $V_b=2.8$~V, fragile superconductivity forms adjacent to the resistive bump. The two shift leftward with increasing $B_{\parallel}$, and the superconducting pocket is seen to extend over a range of $\approx 1.5$~T in $B_{\parallel}$ (Fig.~\ref{fig:4}e).

A measurement of d$V$/d$I$ versus $I_{dc}$ and $B_{\perp}$ confirms nonlinearities characteristic of superconductivity within that pocket (Fig.~\ref{fig:4}f). Importantly, the d$V$/d$I$ map also shows nonreciprocal behavior, with slightly different critical currents $I_c^+ \neq I_c^-$ and distinct associated values of d$V$/d$I$ at fixed $B_\perp$, while obeying the time-reversal symmetry constraint $I_c^+(B_\perp)=I_c^-(-B_\perp)$ (Fig.~\ref{fig:4}g). This superconducting diode effect requires broken inversion and time-reversal symmetries, as would be expected if the parent state is valley-imbalanced. The absence of magnetic hysteresis in the superconducting response (as reported in cases with apparently full valley polarization~\cite{Han2024, Morisette2025SC}) suggests that any valley imbalance is small or forms only short-range domains. We note that diodic behavior is allowed by symmetry in a magnetic field owing to explicit breaking of time reversal. However, nonreciprocal transport appears only for a subset of gate voltages along the river (Extended Data Fig.~\ref{fig:DiodeEffect}), suggesting the presence of different underlying orders along the ridge.

As $D$ is increased further, up and to the left in Fig.~\ref{fig:4}a from the gate voltages where superconductivity appears, the system undergoes an abrupt transition to an integer quantum anomalous Hall (IQAH) state, very likely corresponding to complete spin and valley polarization. The IQAH state is robust to in-plane field, with near-zero $\rho_{xx}$ and $\rho_{xy}$ approaching $h/e^2$ even at $B_{\parallel}=4$~T (Figs.~\ref{fig:4}b–d). Interestingly, the data at 4~T reflect a reversed sign of the Chern number compared to those at $B_\parallel=0$ (see Extended Data Fig.~\ref{fig:IQAH0Tvs4T}), similar to recent reports on moir\'e R6G~\cite{Xie2025_2} and twisted bilayer-trilayer graphene~\cite{Su2025}. Although the superconductivity approaches the IQAH region, the two phases do not directly connect. Because valley imbalance implies a finite center-of-mass momentum for Cooper pairs, it is generally pair-breaking and compatible with superconductivity only in special cases~\cite{Han2024,Morisette2025SC}. 

\medskip\noindent\textbf{Discussion}

Several factors point to the superconductivity we observe being spin triplet. First, it is induced by in-plane magnetic fields over a wide range of the $(V_b,V_t)$ plane and can persist to $B_{\parallel}>6$~T (Extended Data Fig.~\ref{fig:BparVtMaps}). This can be compared to the Pauli paramagnetic limit, which predicts pair breaking when the Zeeman energy exceeds the superconducting gap. For a singlet superconductor with a g-factor of 2, the Pauli limit corresponds to a maximal in-plane upper critical field of $B_{P}\approx 1.25 \frac{k_B T_c}{\mu_B}$, where $k_B$ is the Boltzmann constant and $\mu_B$ is the Bohr magneton. We find that this limit ($B_{P} \approx 0.2$~T for $T_c = 100$~mK) is strongly violated at all positions along the superconducting river that we examined, often by more than an order of magnitude (Extended data Fig.~\ref{fig:PVRlimit}d). The Pauli limit is ill-defined for field-induced superconductivity (since $T_c=0$ at $B=0$), yet we still find strong violations when taking the maximal value of $T_c$ at any $B_{\parallel}$ (see Methods for details). Second, it appears at the edge of a phase that is apparently spin-polarized, as evidenced by the leftward or downward shift of the phase boundary within the $(V_b,V_t)$ plane as $B_{\parallel}$ is raised, lowering the energy of the spin-polarized phase on the right relative to the unpolarized phase on the left. Together with evidence for an adjacent spin-polarized phase, these large Pauli-limit violations are most consistent with a spin-triplet (valley-singlet) order parameter.

The evolution of superconductivity with $B_{\parallel}$ depends strongly on position along the river in the $(V_b,V_t)$ plane. In the band-overlapping regime, where superconductivity generally follows a contour of fixed $V_t$, the Fermi level likely remains near constant filling of the conduction band. This implies that factors beyond the conduction band Fermi surface govern how the pairing strength evolves with gating. In particular, evolution of the valence band Fermi surface with $V_b$ can alter screening of long-range Coulomb interactions, affecting the balance with Hund's coupling, spin-orbit coupling, and pairing instabilities. Consistent with this picture, we observe abrupt changes in the QOs over the same range of $V_b$ where SC1 appears (Extended Data Fig.~\ref{fig:ConstantVt_FFT}), pointing to a reconstruction of the valence band on the bottom crystal surface that modifies screening of the conduction band on the top. Because superconductivity survives at $B_{\parallel}$ far beyond the Pauli limit, and an in-plane field is not expected to be pair breaking for spin-triplet superconductivity, its eventual suppression with $B_{\parallel}$ is most naturally ascribed to orbital depairing effects~\cite{Holleis2025}. Further work is needed to establish a quantitative connection between the band-structure evolution with $(V_b,V_t)$, in-plane orbital magnetism, and pair breaking.

The intrinsic (Kane-Mele) spin–orbit coupling of graphene~\cite{KaneMele2005Quantum}, though relatively small ($\approx25$ to $100~\mu$eV~\cite{Arp2024,Sichau2019,Banszerus2020,Kurzmann2021}), may still play an important role for superconductivity due to the small pairing gap, $\Delta \sim 15~\mu$eV for a BCS state with $T_c \sim 100$~mK. When projected to a predominantly layer-polarized band, Kane--Mele SOC acts like an Ising SOC term and favors oppositely oriented out-of-plane spins locked to each valley. By contrast, Hund’s interaction favors collinear spin alignment~\cite{Chatterjee2022intervalley}. When the density of states is sufficiently large to promote symmetry breaking, the resulting spin-polarized phase may adopt either an out-of-plane (easy-axis) or in-plane (easy-plane) configuration, depending on the relative strengths of these anisotropies and Coulomb interactions (see Supplementary Information for a minimal symmetry-informed model that explores this interplay). In the easy-axis spin-polarized configuration, one spin necessarily anti-aligns with the direction preferred by its SOC-determined spin-valley locking. As a result, the associated valley depopulates to minimize the density of disfavored spins, simultaneously enlarging the favored valley and producing a net valley imbalance. This configuration likely suppresses intervalley pairing, since Cooper pairs would acquire a nonzero center-of-mass momentum. Upon applying $B_{\parallel}$, the spins gradually rotate into the plane and valley balance is restored (Fig.~\ref{fig:3}a inset, and Supplementary Information). 

A possible scenario consistent with our observations is that the spin polarization points predominantly out of plane along most of the resistive stripe, except near SC1 and possibly SC2. In this picture, the spin-polarized phase to the right is in the canted (easy-plane) state, and a narrow valley-imbalanced phase (e.g., a partially-valley-polarized quarter- or three-quarters metal) emerges at the boundary with the unpolarized metal to the left. A small $B_{\parallel}$ rotates out-of-plane spins into the plane, reducing the valley imbalance and enabling superconductivity to form along the entire stripe. This mechanism also plausibly explains gradual enhancement of $T_c$ with $B_{\parallel}$. By contrast, the easy-plane spin-polarized state is valley balanced at $B_\parallel=0$ ~\cite{Patterson2025,Koh2024trilayer,Dong2025superconductivity}. Increasing $B_{\parallel}$ simply decreases the tiny out-of-plane canting angle induced by Kane--Mele SOC and orbital depairing should immediately reduce $T_c$. As noted earlier, the formation of valley-imbalanced domains can also account for the observed nonlinear $\mathrm{d}V/\mathrm{d}I$ at small $B_\parallel$ (Fig.~\ref{fig:2}e), although other explanations remain possible. 

If superconductivity can form despite a small residual valley imbalance, the parent state simultaneously breaks inversion and time-reversal symmetries, naturally producing the superconducting diode effect observed in Figs.~\ref{fig:4}f–g and Extended Data Fig.~\ref{fig:DiodeEffect}. The absence of a diode effect in SC1 (Extended Data Fig.~\ref{fig:DiodeEffect}f,g) is then consistent with a valley-balanced, spin-canted state at $B_\parallel=0$. We caution that further work is needed to test these ideas, including direct characterization of the parent-state orbital magnetization with SQUID magnetometry~\cite{Patterson2025, Arp2024}, compressibility mapping of thermodynamic phase boundaries versus $B_{\perp}$ and $B_{\parallel}$~\cite{Arp2024}, and experiments with proximal TMD layers to directly tune the strength of SOC~\cite{Zhang2023,Zhang2025,Yang2025,Seo2025_UncSC,Yang2025_MagSC,Li2024,Holleis2025, Patterson2025}. It is also possible that the resistive stripe at $B=0$ is intervalley-coherent or nematic in addition to being spin polarized, although our transport measurements are not directly sensitive to these orders. 

Given that several features of our results also appear in BBG and rhombohedral trilayer graphene, we speculate that spin-triplet superconductivity may be generic across rhombohedral graphene over a wide range of layer numbers and does not require a moir\'e potential. Nevertheless, the moir\'e potential has at least two clear effects in our measurements. First, it suppresses the correlated state at $n=D=0$ seen in misaligned devices and proposed to be a spontaneous layer-polarized phase~\cite{Nilsson2006,Zhang2011,Velasco2012,Han2024Nanotech,Liu2024,Han2023Nature,Myhro2018}. This correlated phase typically interrupts the sharp conduction band resistive peak that forms the superconducting river in our sample; in its absence, we instead observe a single extended superconducting feature. Second, a moir\'e pattern is required to realize integer and fractional quantum anomalous Hall states. In our device it enables superconductivity to appear in proximity to a $C=1$ Chern insulator. The close adjacency of these phases motivates future experiments that interface them electrostatically, with the goal of creating non-abelian anyons.


\section*{Methods}

\textbf{Device fabrication.} The moir\'e R7G device here was previously reported in Ref.~\cite{Kumar2025_dual}. Fabrication details can be found in that work. In short, we first identified and isolated a rhombohedral graphene domain~\cite{Waters2025,Li2018} with the layer number determined optically~\cite{Kumar2025_dual}, and then used standard dry transfer techniques with a polycarbonate (PC) film on a polydimethylsiloxane (PDMS) stamp for vdW assembly. From the top down, the device consists of graphite, hBN, rhombohedral graphene, hBN, and graphite, all resting on a Si/SiO$_2$ wafer. Standard device fabrication procedures were employed to create the dual-gated Hall bar device (i.e., reactive ion etching and evaporation of 7/70 nm of Cr/Au, all using poly(meth)acrylate (PMMA) masks patterned by e-beam lithography).

\textbf{Transport measurements.} Transport measurements were carried out in a Bluefors XLD dilution refrigerator with a one-axis superconducting magnet as well as a Bluefors LD dilution refrigerator equipped with a 3-axis superconducting vector magnet. Unless otherwise specified, measurements were carried out at the nominal base mixing chamber temperature of the fridge ($T=8$~mK, as measured by a factory-supplied RuO$_x$ sensor). The cryostat temperature sensor occasionally produced spurious zero readings. These data points were identified and replaced by interpolated values, calculated as the arithmetic mean of the nearest valid measurements before and after the zero reading. 

Four-terminal lock-in measurements were performed by sourcing a small alternating current between $I_{ac}=1$~nA and $10$~nA at a frequency $<40$ Hz, chosen to accurately capture sensitive transport features while minimizing electronic noise. Resistivity values can have a systematic uncertainty of a few percent due to the use of different voltage divider setups. The current was calculated more precisely for measurements of the IQAH state in order to minimize this error. All $\rho_{xx}$ and d$V$/d$I$ measurements are presented after multiplying the raw measured resistance values by the aspect ratio $W/L$, set by the width ($W$) and length ($L$) of the portion of the Hall bar between the voltage probes. In addition, a global bottom gate voltage between $-20$~V and $+20$~V was applied to the Si substrate to improve the contact resistance. Unless otherwise noted, all $\rho_{xx}$ data corresponds to measurements on the voltage probes labeled as $V_{xx1}$ in Extended Data Fig.~\ref{fig:Optical micrograph and n-D maps}a.

The charge carrier density, $n$, and the out-of-plane electric displacement field, $D$, were defined according to $n= \left(C_{\text{b}} V_{\text{b}}+C_{\text{t}} V_{\text{t}}\right) / e$ and $D=\left(C_{\text{b}} V_{\text{b}} - C_{\text{t}} V_{\text{t}}\right) / 2 \epsilon_0$, where $C_{\text{t}}$ and $C_{\text{b}}$ are the top and bottom gate capacitance per unit area and $\epsilon_0$ is the vacuum permittivity. $C_{\rm{t}}$ and $C_{\rm{b}}$ were estimated by fitting the slopes of the quantum Hall states in Landau fan diagrams. The moir\'e period of 13.5~nm was also extracted from Landau fan diagrams following the process described in Ref.~\cite{Kumar2025_dual}.

In some Landau fan diagrams, we symmetrized $\rho_{xx}$ and antisymmetrized $\rho_{xy}$ to reduce the effects of geometric mixing, following the relations $\rho_{xx}=\left(\rho_{xx}(B > 0) + \rho_{xx}(B < 0)\right)/2$ and $\rho_{xy}=\left(\rho_{xy}(B > 0) - \rho_{xy}(B < 0)\right)/2$. In measurements of superconductivity versus $B$, we adjusted the nominal value of $B$ by a few millitesla such that the $\rho_{xx}$ data is symmetric about $B=0$, which was necessary due to the small trapped flux in the superconducting magnet coil.

We report critical temperature ($T_c$) and critical in-plane field ($B_{\parallel{c}}$) in Fig.~\ref{fig:3}, Extended Data Fig.~\ref{fig:PVRlimit}, and Extended Data Fig.~\ref{fig:TC_Vb_B}. We estimate these values by fitting a line to the normal state of the measured $\rho_{xx}$ curve then determining the temperature or field at which $\rho_{xx}$ intersects with the same line multiplied by 0.9. In effect, this sets the critical temperature and critical field to where the resistance drops to 90\% of its normal state value.  

\textbf{Calibrating the vector field orientation.} The superconducting states in this device are extremely sensitive to $B_\perp$, but remain relatively robust against $B_\parallel$. For all measurements involving the application of an in-plane magnetic field, we thus carefully calibrated the vector magnet to eliminate any residual out-of-plane component. For each in-plane field setting, we performed a $B_\perp$ versus $I_\mathrm{dc}$ measurement and identified the value of $B_\perp$ at which the critical current $I_c$ was maximized. This value was then taken as the properly aligned in-plane field.

\textbf{Non-zero resistance saturation and Fraunhofer oscillations.} 
An unexpected feature of the superconductivity is that the resistance often saturates to a small, nonzero value at base temperature. Similar behavior is common in rhombohedral graphene devices---especially for thicker stacks---though its origin remains unclear~\cite{Zhou2021_RTGSC, Patterson2025, Holleis2025, Choi2025, Zhang2025, Yang2025, Seo2025_UncSC, Deng2025_Xiaomeng}. In our sample, the saturation value depends on position along the river in the $(V_b,V_t)$ plane and on $B_{\parallel}$. As these parameters are tuned, the saturated resistance ranges from $\sim$\,220~$\Omega$ [$V_b=3.10$~V, $V_t=-3.70$~V, $B_{\parallel}=0.5$~T] down to $\sim$\,7~$\Omega$ [$V_b=4.75$~V, $V_t=-2.00$~V, $B_{\parallel}=2$~T] (Supplementary Information Fig.~\ref{fig:RscRnorm}).

Despite the nonzero $R_{\mathrm{sat}}$, we observe clear Fraunhofer oscillations in $\mathrm{d}V/\mathrm{d}I$ maps versus $I_{\mathrm{dc}}$ and $B_{\perp}$ across many gate settings (Extended Data Fig.~\ref{fig:Fraunhoffer}), which are hallmarks of phase-coherent Josephson transport. The oscillation period $\Delta B$ sets an effective junction area via $A_{\mathrm{eff}}=\Phi_0/\Delta B\approx1.6$~${\mu}m^2$, and the lobe structure of $I_c(B_{\perp})$ is consistent with interference through a spatially extended weak link. Together, these features establish superconducting phase coherence even when a small residual resistance is present.

\textbf{Fermiology Analysis.} Fast Fourier transforms (FFTs) of $\rho_{xx}$($\frac{1}{B}$) are taken for the Landau fans shown in Extended Data Fig.~\ref{fig:PRB_FFT} and Extended Data Fig.~\ref{fig:ConstantVt_FFT}. Before computing the FFTs, we first subtract a fifth-order polynomial from the raw $\rho_{xx}$ data, following Ref.~\onlinecite{Zhou2021_2}. We then interpolate the subtracted data onto a regular grid in order to take the FFT. For Extended Data Fig.~\ref{fig:PRB_FFT}b the FFT is taken over a magnetic field range of $0.09-0.93$~T, for Extended Data Fig.~\ref{fig:PRB_FFT}c over a range of $0.07-1$~T, for Extended Data Fig.~\ref{fig:PRB_FFT}d over a range of $0.08-1$~T, for Extended Data Fig.~\ref{fig:PRB_FFT}e over a range of $0.08-0.52$~T, and for Extended Data Fig.~\ref{fig:PRB_FFT}f over a range of $0.11-0.56$~T. For Extended Data Fig.~\ref{fig:ConstantVt_FFT} the FFT is taken over a magnetic field range of $0.08-0.80$~T.

After taking the FFT, we normalize the raw frequencies by the Luttinger volume, $n(\frac{h}{e})$. Normalized frequencies ($f_v$) correspond to the fraction of the total fermi surface enclosed by a cyclotron orbit in momentum space. This implies a sum rule, wherein for a given carrier density the sum of all frequencies adds to one (accounting for carrier sign, isospin degeneracies, and the Fermi surface topology).

The code we use for analyzing the FFT is based on the code provided in Ref.~\onlinecite{Zhang2023} and was initially developed for use in Ref.~\onlinecite{Kumar2025_dual}. The procedure used to fit frequencies is outlined in detail in Ref.~\onlinecite{Kumar2025_dual}. Briefly, the normalized FFT is cut-off at a maximum frequency and interpolated onto a rectangular grid, which we further make into an interactive grid. The user then directly selects individual trajectories in the FFT. User-selected arrays are then fit to polynomials. We use the following line-style conventions in our FFT analysis plots: solid lines are user-selected trajectories as described above, or are constant fractions (e.g. $\frac{1}{4}$); dashed lines are derived from solid lines, either by adding/subtracting a fraction, or by multiplying by an integer. We also use the following color conventions in the FFT analysis plots: blue corresponds to lines (or groups of lines) which sum to $\frac{1}{2}$, red corresponds to $\frac{1}{4}$ and green to $\frac{1}{12}$. All other lines are colored black. The shown fits represent our best understanding of the FFTs, however, we note that other interpretations may be possible.

\textbf{Phenomenological framework: valley imbalance and field response.}
We consider two scenarios for a spin-polarized half-metal phase once intrinsic spin–orbit coupling ($\lesssim 0.1$~meV) is included. \emph{Scenario~1} (easy axis): spins polarize out of the plane. This state is valley imbalanced due to the spin polarization acting in conjunction with or against Kane-Mele SOC, depending on the valley (see top half of the inset of Fig.~\ref{fig:3}a). \emph{Scenario~2} (easy plane): spins lie primarily in the 2D plane with a small out-of-plane canting, in opposite directions in the two valleys, due to Kane–Mele SOC. This state is valley balanced by symmetry~\cite{Patterson2025,Koh2024trilayer,Dong2025superconductivity} (see bottom half of the inset of Fig.~\ref{fig:3}a). 

The energetic competition between the two scenarios is expected to be quite subtle and to depend on details of the band structure as well as the relative strengths of long-range Coulomb interactions, Hund's coupling, and Kane-Mele SOC. In the Supplementary Information we present a phenomenological model exploring this physics. We find that, in a spin-polarized half metal, Kane-Mele SOC is most susceptible to induce a valley imbalance when the system is close to transitioning to a quarter-metal or three-quarters metal.

Our working hypothesis is that the sharp resistive ridge corresponds to \emph{Scenario~1}, bounded on its left by an unpolarized (or weakly spin-polarized) phase and on its right by a (fully or partially polarized) spin-canted phase corresponding to \emph{Scenario~2} (Fig.~\ref{fig:3}a). It also remains possible that the underlying physics is described by a mechanism we have not yet considered---for example, relying on an interplay with intervalley coherence or nematicity.

The following are predictions and consistency checks for this hypothesis, some of which are preliminarily addressed in our work and others left for future studies:

(i) \emph{Nonlinear transport on the ridge.} A valley-imbalanced state can host orbital-magnetization domains whose domain walls are driven by $I_{\mathrm{dc}}$~\cite{Tschirhart2023SpinHallTorque}, producing pronounced nonlinear $\mathrm{d}V/\mathrm{d}I$ features on the ridge, while adjacent (spin-canted) metallic regions remain nearly linear.  

(ii) \emph{Effect of in-plane field.} In this scenario, $B_{\parallel}$ rotates spins into the plane and reduces valley imbalance. Because valley imbalance is pair-breaking for intervalley Cooper pairs, diminishing it (a) enables \emph{field-induced} superconductivity when the zero-field parent state is metallic and (b) yields a \emph{non-monotonic} $T_c(B_{\parallel})$: $T_c$ rises as spins planarize and pair-breaking weakens, then falls at larger $B_{\parallel}$ due to in-plane orbital depairing.

(iii) \emph{Superconducting diode effect.} When weak valley imbalance coexists with superconductivity, the breaking of inversion and time-reversal symmetries allows for non-reciprocal transport. The magnitude of the superconducting diode effect should be largest at the lowest $B_{\parallel}$ and diminish as spins fully align in plane.  

(iv) \emph{Local exceptions.} SC1, which is strongest at $B=0$, is expected to be valley balanced and therefore should not exhibit a diode effect at any $B_{\parallel}$. SC2 may sit closer to the valley-imbalanced regime: it can exist at $B=0$ yet share the non-monotonic $T_c(B_{\parallel})$ seen along the field-induced segments. 

(v) \emph{Fermiology signature.} Valley-imbalanced phases should show a weak splitting of the primary quantum-oscillation frequency. We are not able to resolve this feature experimentally owing to the narrowness of the resistive bump feature with gate voltage.

(vi) \emph{Signatures in compressibility.} Putative phase transitions between the valley-imbalanced phase and either an unpolarized or spin-canted phase could also be studied using compressibility measurements as a function of in-plane and out-of-plane magnetic fields~\cite{Arp2024,Patterson2025}. The valley-imbalanced phase is characterized by a spontaneously generated out-of-plane magnetic moment (consisting of an orbital and spin contribution) but no in-plane moment; the spin-canted phase sits in the opposite limit with only the in-plane moment non-zero. Therefore, for small applied fields the valley-imbalanced phase should expand with $B_\perp$ against both of its neighbors. In contrast, applying weak $B_\parallel$ should leave the transition to an unpolarized phase unchanged but favor spin canting over valley imbalance.

\section*{Acknowledgments}

The authors thank UBC research associate Silvia Folk for helpful contributions, and Andrea Young and Cyprian Lewandowski for helpful discussions. Research on superconductivity was supported by the Army Research Office under award number W911NF-25-1-0012. Work on topology was supported as part of Programmable Quantum Materials, an Energy Frontier Research Center funded by the U.S. Department of Energy (DOE), Office of Science, Basic Energy Sciences (BES), under award DE-SC0019443. Device fabrication was supported by National Science Foundation (NSF) CAREER award no. DMR-2041972 and University of Washington Molecular Engineering Materials Center, a U.S. National Science Foundation Materials Research Science and Engineering Center (DMR-2308979). Experiments at the University of British Columbia were undertaken with support from the Natural Sciences and Engineering Research Council of Canada; the Canada Foundation for Innovation; the Canadian Institute for Advanced Research; the Max Planck-UBC-UTokyo Centre for Quantum Materials and the Canada First Research Excellence Fund, Quantum Materials and Future Technologies Program; and the European Research Council (ERC) under the European Union’s Horizon 2020 research and innovation program, Grant Agreement No. 951541. A.O. and M.Y. acknowledge support from the State of Washington-funded Clean Energy Institute. E.L.-H is grateful for support from the National Sciences and Engineering Council of Canada (NSERC), grant RGPIN-2025-06136, the Regroupement québécois sur les matériaux de pointe (RQMP), and start-up funds from the Faculté des Sciences at Université de Sherbrooke. K.W. and T.T. acknowledge support from the JSPS KAKENHI (Grant Numbers 21H05233 and 23H02052) and World Premier International Research Center Initiative (WPI), MEXT, Japan. This work made use of shared fabrication facilities at UW provided by NSF MRSEC 2308979.

\section{Author Contributions} 
M.K. fabricated the samples with assistance from A.O. M.K., D.W. and A.O. led measurements at UBC remotely with R.T. assisting on-site. E.L.-H. performed the theoretical modeling. K.W. and T.T. provided the hBN crystals. J.F. and M.Y. supervised the project. 

\section*{Competing interests}
The authors declare no competing interests.

\section*{Additional Information}
Correspondence and requests for materials should be addressed to Joshua Folk or Matthew Yankowitz.

\section*{Data Availability}
Source data are available for this paper. All other data that support the findings of this study are available from the corresponding author upon request.

\bibliographystyle{naturemag}
\bibliography{references}

\newpage

\renewcommand{\figurename}{Extended Data Fig.}
\renewcommand{\thesubsection}{S\arabic{subsection}}
\setcounter{secnumdepth}{2}
\renewcommand{\thetable}{S\arabic{table}}
\setcounter{figure}{0} 
\setcounter{equation}{0}

\onecolumngrid
\newpage

\FloatBarrier
\section*{Extended Data}

\begin{figure*}[h]
\includegraphics[width=0.95\textwidth]{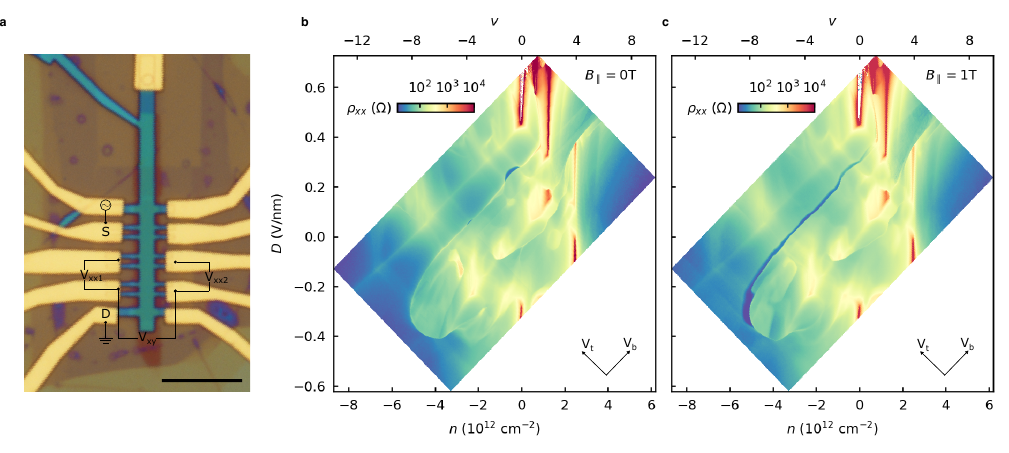} 
\caption{\textbf{Device image and $n-D$ maps.} \textbf{a}, Optical micrograph of the R7G sample with source, drain and voltage probe contacts annotated. The scale bar is 10 $\mu$m. \textbf{b}, Map of $\rho_{xx}$ plotted as a function of $n-D$ axis (see Methods for conversion), taken at $B_{\parallel}=0$~T. \textbf{c}, same as (\textbf{b}) but taken at $B_{\parallel}=1$~T. 
}
\label{fig:Optical micrograph and n-D maps}
\end{figure*}

\begin{figure*}[h]
\includegraphics[width=0.85\textwidth]{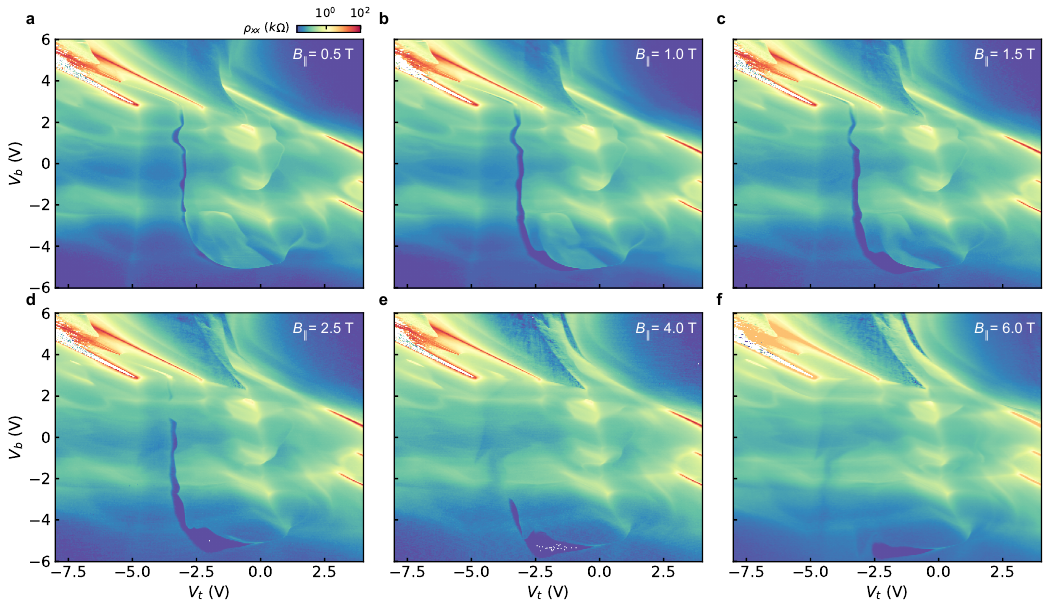} 
\caption{\textbf{Extent of superconductivity at different $B_{\parallel}$.} Maps of $\rho_{xx}$ versus $V_b$ and $V_t$ taken at \textbf{a}, $B_\parallel=0.5$~T, \textbf{b}, $1$~T, \textbf{c}, $1.5$~T, \textbf{d}, $2.5$~T, \textbf{e}, $4$~T, and \textbf{f}, $6$~T.
}
\label{fig:FiniteFieldMaps}
\end{figure*}

\begin{figure*}[t]
\includegraphics[width=0.95\textwidth]{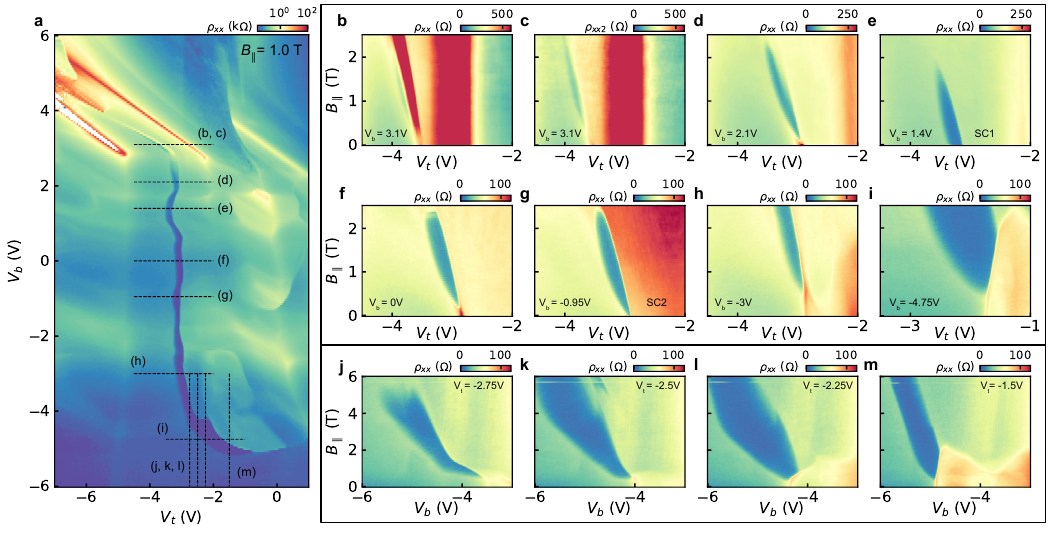} 
\caption{\textbf{Evolution of superconductivity with $B_{\parallel}$.} \textbf{a}, Map of $\rho_{xx}$ versus $V_b$ and $V_t$ taken at $B_\parallel=1$~T. \textbf{b}, $\rho_{xx}$ versus $B_\parallel$ and $V_t$ taken at $V_b=3.1$~V (marked by the black dashed line in \textbf{(a)}) using $V_{xx1}$. \textbf{c}, Same as \textbf{(b)} using $V_{xx2}$. \textbf{d}, Same as \textbf{(b)} taken at $V_b=2.1$~V, \textbf{e}, $V_b=1.4$~V, \textbf{f}, $V_b=0$~V, \textbf{g}, $V_b=-0.95$~V, \textbf{h}, $V_b=-3$~V, and \textbf{i}, $V_b=-4.75$~V. \textbf{j}, $\rho_{xx}$ versus $B_\parallel$ and $V_b$ taken at $V_t=-2.75$~V. \textbf{k}, Same as \textbf{(j)} taken at $V_t=-2.5$~V, \textbf{l}, $V_t=-2.25$~V, and \textbf{m}, $V_t=-1.5$~V. Unless otherwise specified, measurements are taken using $V_{xx1}$. Boxes around \textbf{(b-i)} and \textbf{(j-m)} delineate maps taken at fixed $V_b$ and at fixed $V_t$ respectively.
}
\label{fig:BparVtMaps}
\end{figure*}

\begin{figure*}[t]
\includegraphics[width=0.7\textwidth]{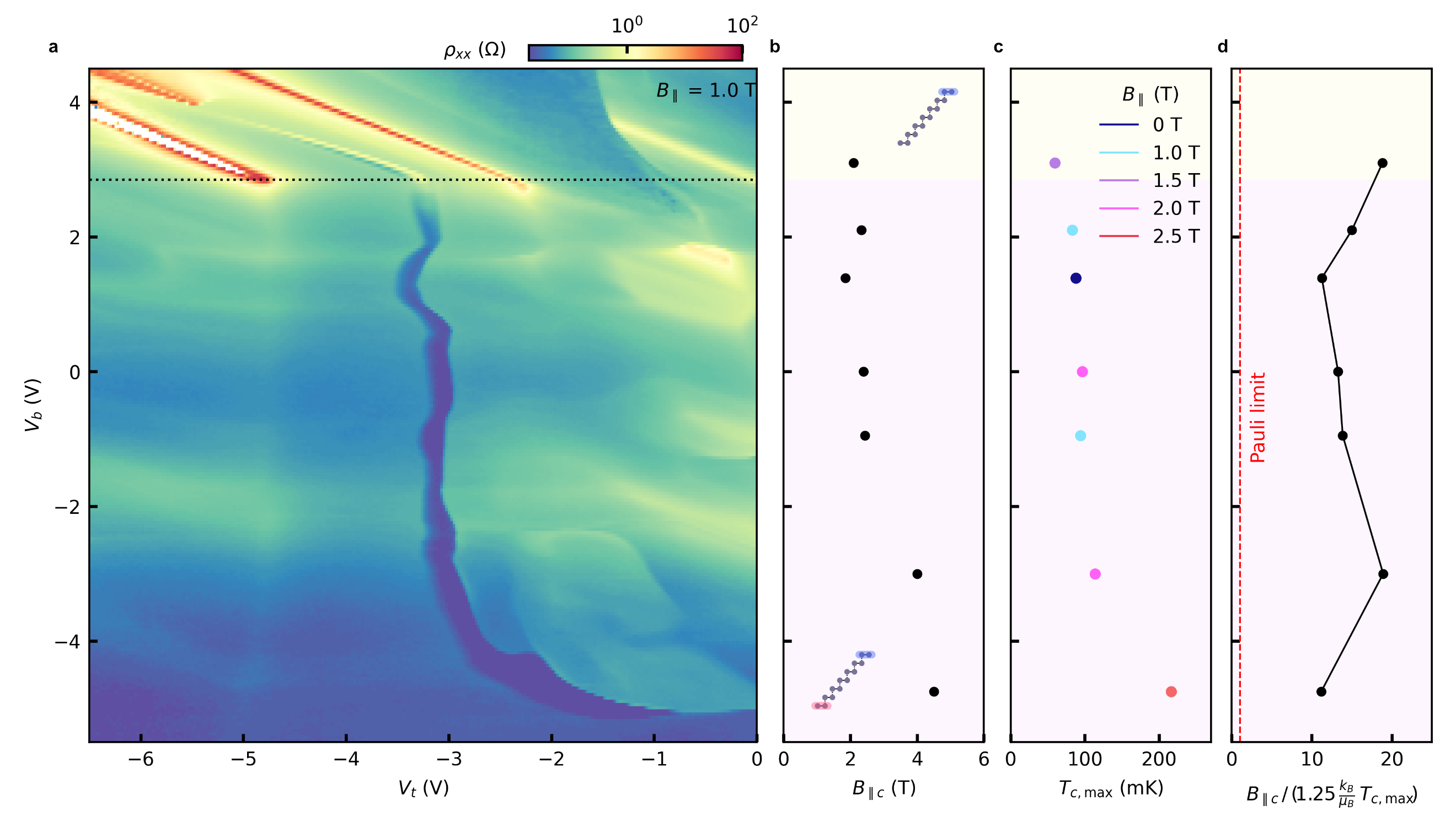} 
\caption{\textbf{Assessment of the Pauli limit violation.} \textbf{a}, Map of $\rho_{xx}$ versus $V_b$ and $V_t$ taken at $B_\parallel$ = 1 T. The dashed line shows the boundary between band isolated (above) and band overlap (below) regimes. \textbf{b}, Critical in-plane field, $B_{\parallel{c}}$, in the superconducting river as a function of $V_b$. The values of $B_{\parallel{c}}$ at $V_b$ = -3, -4.75 V are an estimated lower bound. The values of $B_{\parallel{c}}$ at $V_b$ = 3.1, 2.1, 1.4, 0, -0.95 V are extracted by taking 90\% of the normal state resistance. The lattice schematics show the polarization of the wavefunction to either the top layer of the rhombohedral graphene in the yellow band isolated regime, or to both the top and bottom layers in the purple band overlap regime. \textbf{c}, Plot of the maximum critical temperature for all measured values of $B_\parallel$, denoted $T_{c, max}$, as a function of $V_b$. The legend shows at which $B_\parallel$ each value of $V_b$ has its maximal $T_c$. \textbf{d}, Lower bound estimate of the Pauli limit violation as a function of $V_b$, as calculated by ${B_{\parallel{c}}} / ({1.25 \frac{k_B T_{c,max}}{\mu_B})}$. The vertical red dashed lines denotes a Pauli violation ratio of 1, showing that the superconductivity far exceeds the Pauli limit everywhere along the river.}
\label{fig:PVRlimit}
\end{figure*}

\begin{figure*}[t]
\includegraphics[width=0.8\textwidth]{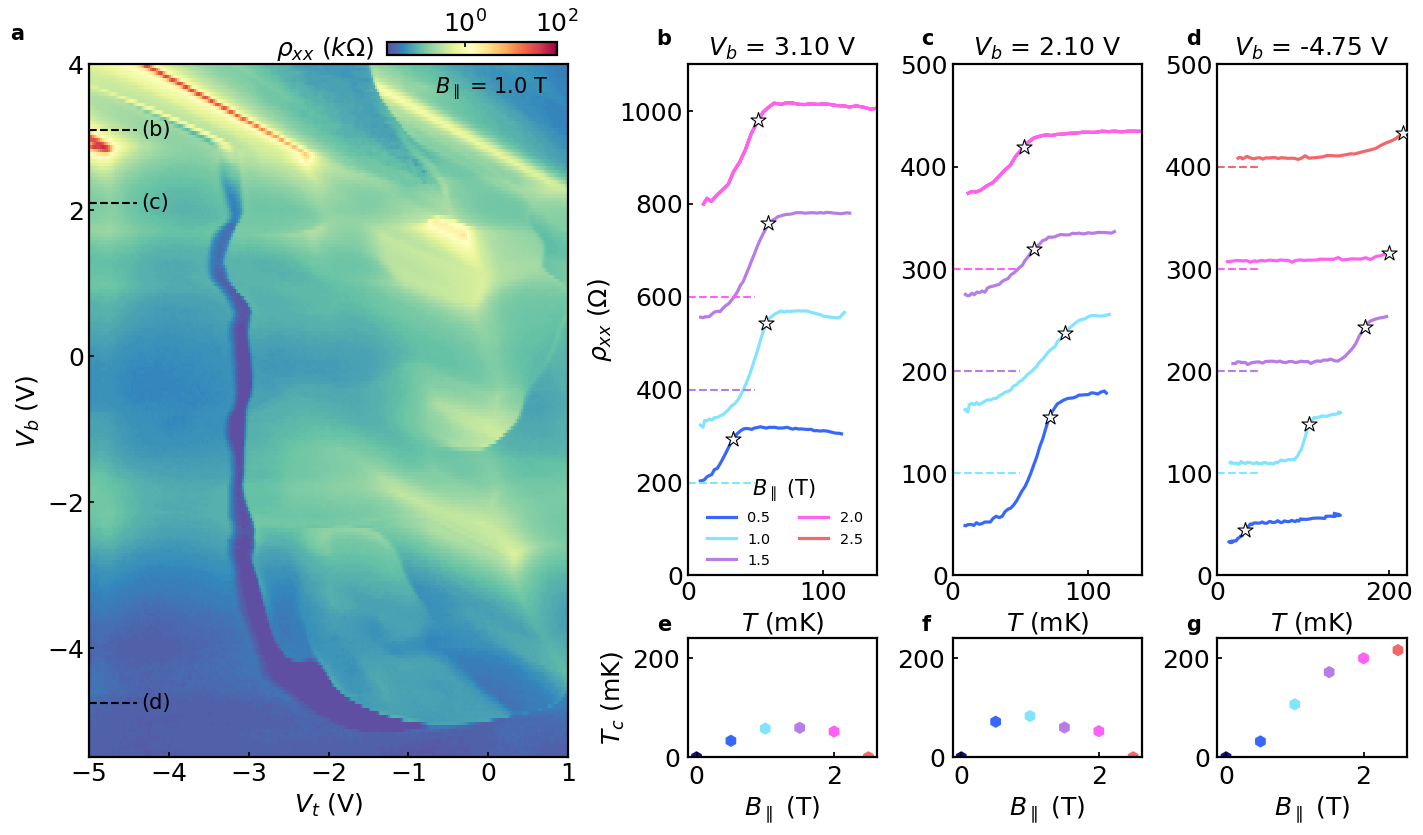} 
\caption{\textbf{Additional measurements showing a non-monotonic evolution of $T_c$ with $B_\parallel$}. \textbf{a}, Map of $\rho_{xx}$ versus $V_b$ and $V_t$ taken at $B_\parallel= 1.0$~T. The dashed lines correspond to the $V_b$ values in (\textbf{b}), (\textbf{c}), and (\textbf{d}). \textbf{b}, $\rho_{xx}$ versus $T$ measurements at several values of $B_\parallel$ with $B_\perp$ = 0. The measurement is taken with $V_b = 3.10$~V and $V_t\approx -2.0$~V. Curves are vertically offset for clarity. Dashed lines denote $\rho_{xx}$ = 0 for each curve of corresponding color. Stars denote $T_c$. The legend shows the value of $B_\parallel$ at which each curve was taken. \textbf{c}, Same for $V_b= 2.10$~V with $V_t\approx-2.0$~V. \textbf{d}, Same for $V_b =-4.75$~V with $V_t\approx-1.0$~V. \textbf{e-g}, Plots of $T_c$ versus $B_\parallel$ for each of the three associated values of $V_b$. Note that these are the analogous measurements from Fig.~\ref{fig:3} of the main text, shown at additional values of $V_b$.}

\label{fig:TC_Vb_B}
\end{figure*}

\begin{figure*}[t]
\includegraphics[width=0.45\textwidth]{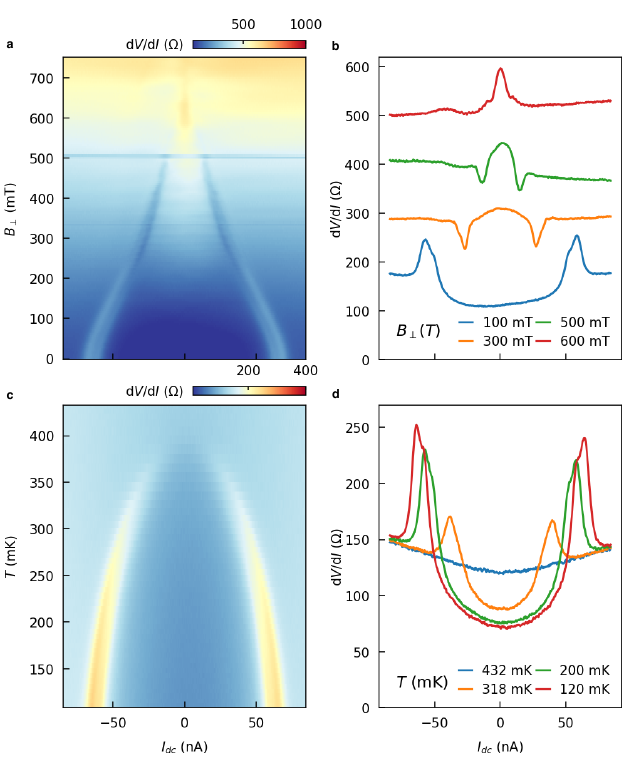} 
\caption{\textbf{Additional characterization of non-linearity associated with the sharp resistive bump.} \textbf{a}, d$V$/d$I$ versus $B_\perp$ and $I_{dc}$ with $B_{\parallel}=0$~T, $V_b=-3$~V and $V_t$ varied with $B_\perp$ to stay on the sharp resistive bump (similar to Fig.~\ref{fig:2}e). \textbf{b}, Line cuts from \textbf{(c)} at fixed values of $B_{\perp}$. \textbf{c}, d$V$/d$I$ versus $T$ and $I_{dc}$ with $B_{\parallel}=0$~T, $V_b=-3$~V and $V_t=-2.75$~V. \textbf{d}, Line cuts from \textbf{(c)} at fixed values of $T$.
}
\label{fig:Non-linearity}
\end{figure*}

\begin{figure*}[t]
\includegraphics[width=0.95\textwidth]{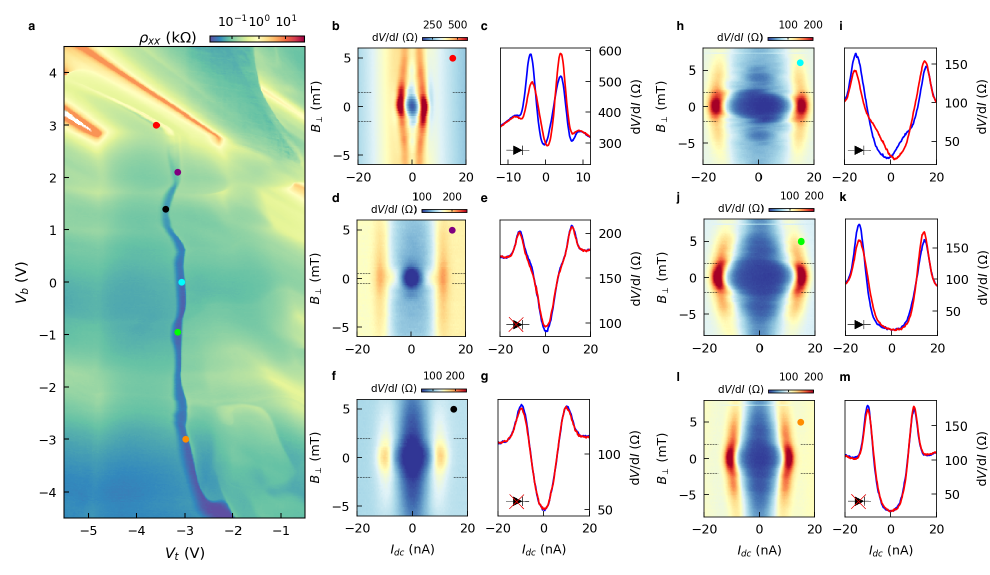}
\caption{\textbf{Gate-tunable nonreciprocity in the superconducting state.} \textbf{a}, Map of $\rho_{xx}$ versus $V_b$ and $V_t$ taken at $B_\parallel=1$~T. Map of d$V$/d$I$ versus $I_{dc}$ and $B_{\perp}$ taken at \textbf{b} ($V_b$,$V_t$)=($3.00$~V$,$$-3.59$~V) (denoted by the red dot in \textbf{(a)}) with $B_{\parallel}=1$~T. \textbf{c}, line cuts from \textbf{(b)} taken at $\pm1.5$~mT (blue and red respectively). \textbf{d}, d$V$/d$I$ map taken at ($2.10$~V, $-3.15$~V) (denoted by the purple dot). \textbf{e}, line cuts from \textbf{(d)} taken at $B_{\perp}=\pm0.5$~mT. \textbf{f}, d$V$/d$I$ map taken at ($1.40$~V, $-3.40$~V) (denoted by the black dot). \textbf{g}, line cuts from \textbf{(f)} taken at $B_{\perp}=\pm2$~mT. \textbf{h},  d$V$/d$I$ map taken at ($0.0$~V, $-3.07$~V) (denoted by the cyan dot). \textbf{i}, line cuts from \textbf{(h)} taken at $B_{\perp}=\pm2$~mT. \textbf{j}, d$V$/d$I$ map taken at ($-0.95$~V, $-3.15$~V) (denoted by the green dot). \textbf{k}, line cuts from \textbf{(j)} taken at $B_{\perp}=\pm2$~mT. \textbf{l}, d$V$/d$I$ map taken at ($-3.00$~V, $-2.98$~V) (denoted by the orange dot). \textbf{m}, line cuts from \textbf{(l)} taken at $B_{\perp}=\pm2$~mT. 
}
\label{fig:DiodeEffect}
\end{figure*}

\begin{figure*}[t]
\includegraphics[width=0.6\textwidth]{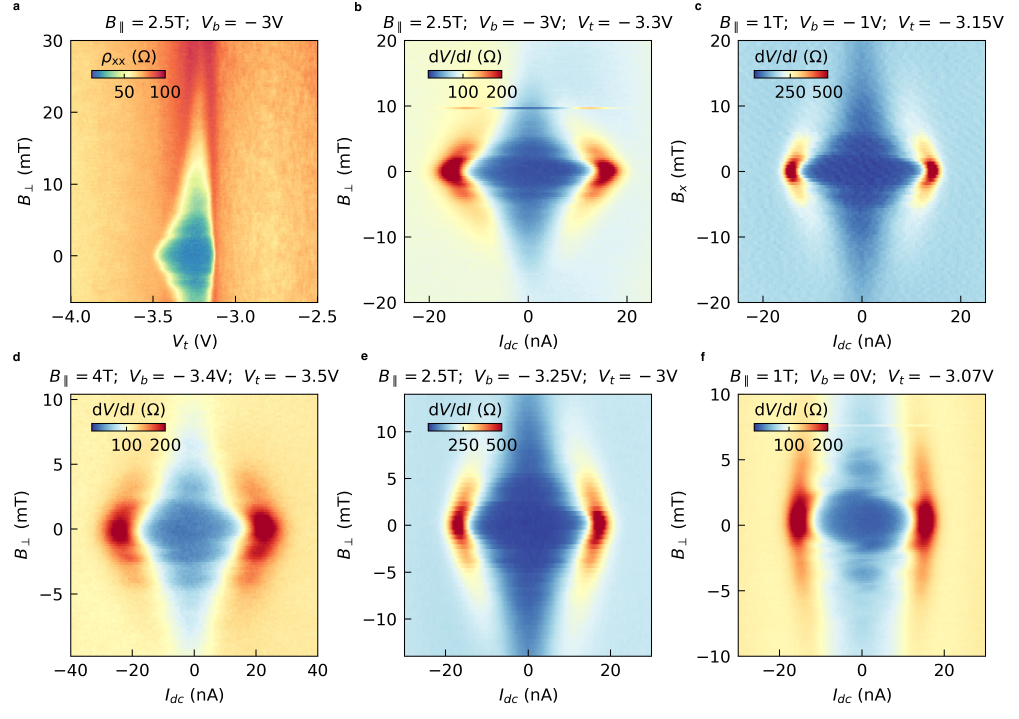}
\caption{\textbf{Examples of Fraunhofer oscillations}. \textbf{a}, Map of $\rho_{xx}$ versus $V_t$ and $B_\perp$ showing Fraunhofer oscillations in the superconducting state, taken at $B_{\parallel}=2.5$~T and $V_b=-3.00$~V. Map of d$V$/d$I$ versus $B_\perp$ and $I_{dc}$ showing Fraunhofer oscillations taken at \textbf{b}, $B_{\parallel}=2.5$~T, $V_b=-3.00$~V and $V_t=-3.30$~V, \textbf{c}, $B_{\parallel}=1$~T, $V_b=-1.00$~V and $V_t=-3.15$~V,  \textbf{d}, $B_{\parallel}=4$~T, $V_b=-3.40$~V and $V_t=-3.50$~V, \textbf{e}, $B_{\parallel}=2.5$~T, $V_b=-3.25$~V and $V_t=-3.00$~, and \textbf{f}, $B_{\parallel}=1.0$~T, $V_b=0.00$~V and $V_t=-3.07$~V.
}
\label{fig:Fraunhoffer}
\end{figure*}

\begin{figure*}[t]
\includegraphics[width=0.95\textwidth]{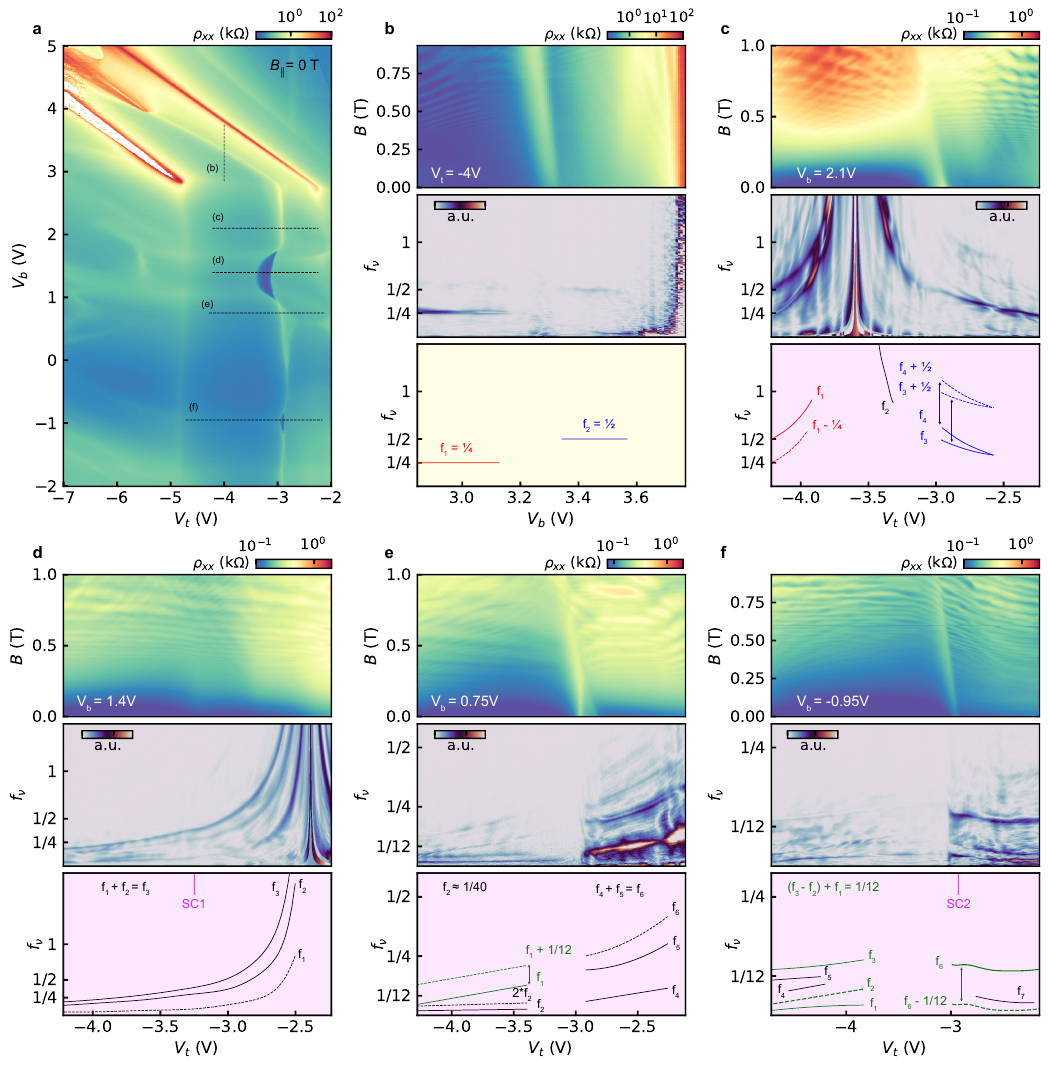}
\caption{\textbf{FFTs crossing the resistive bump feature.} \textbf{a}, Map of $\rho_{xx}$ versus $V_b$ and $V_t$ taken at $B_\parallel=0$~T. \textbf{b}, (top) Landau Fan taken along the dashed black line marked in \textbf{(a)}, at fixed $V_t=-4$~V. (middle) Corresponding FFT. (bottom) Primary frequencies identified in the FFT. \textbf{c}, Same as \textbf{(b)}, for a Landau Fan taken at $V_b=2.1$~V, \textbf{d}, $V_b=1.4$~V, \textbf{e}, $V_b=0.75$~V, and \textbf{f}, $V_b=-0.95$~V. Yellow (purple) backgrounds indicate that the Landau Fan falls within the band-isolated (band-overlap) regime. Purple lines in \textbf{(d)} and \textbf{(f)} mark the positions of SC1 and SC2.
}
\label{fig:PRB_FFT}
\end{figure*}

\begin{figure*}[t]
\includegraphics[width=0.8\textwidth]{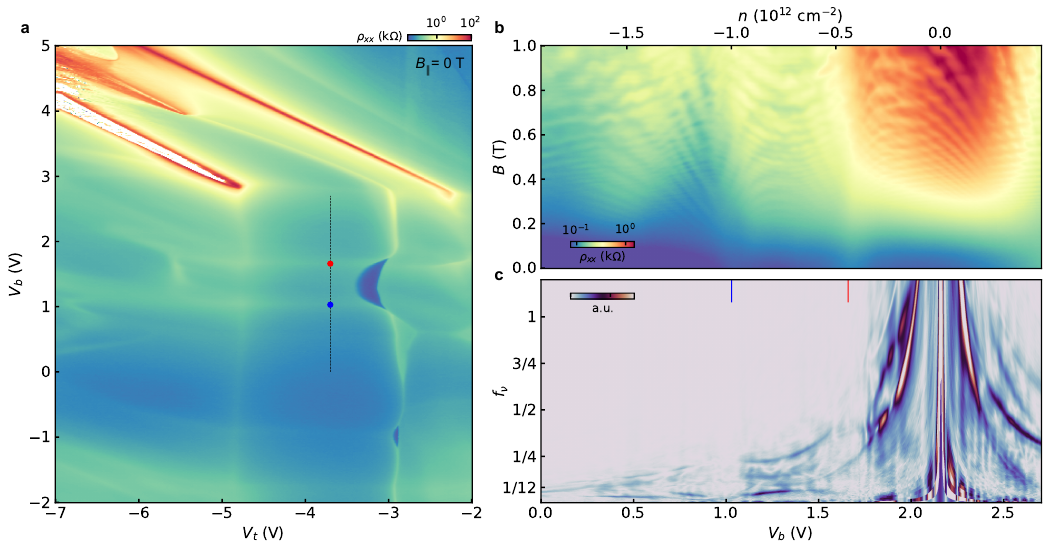}
\caption{\textbf{FFT crossing horizontal resistive features near SC1.} \textbf{a}, Map of $\rho_{xx}$ versus $V_b$ and $V_t$ taken at $B_\parallel=0$~T. \textbf{b}, Landau fan taken along the black dashed line in \textbf{(a)}, at fixed $V_t=-3.7$~V. \textbf{c}, FFT corresponding to the Landau Fan in \textbf{(b)}. Blue and red lines correspond to the blue and red dots in \textbf{(a)}, marking two features which track fixed $V_b$ in \textbf{(a)}. \textbf{(b)} and \textbf{(c)} share their x-axis.
}
\label{fig:ConstantVt_FFT}
\end{figure*}

\begin{figure*}[t]
\includegraphics[width=0.8\textwidth]{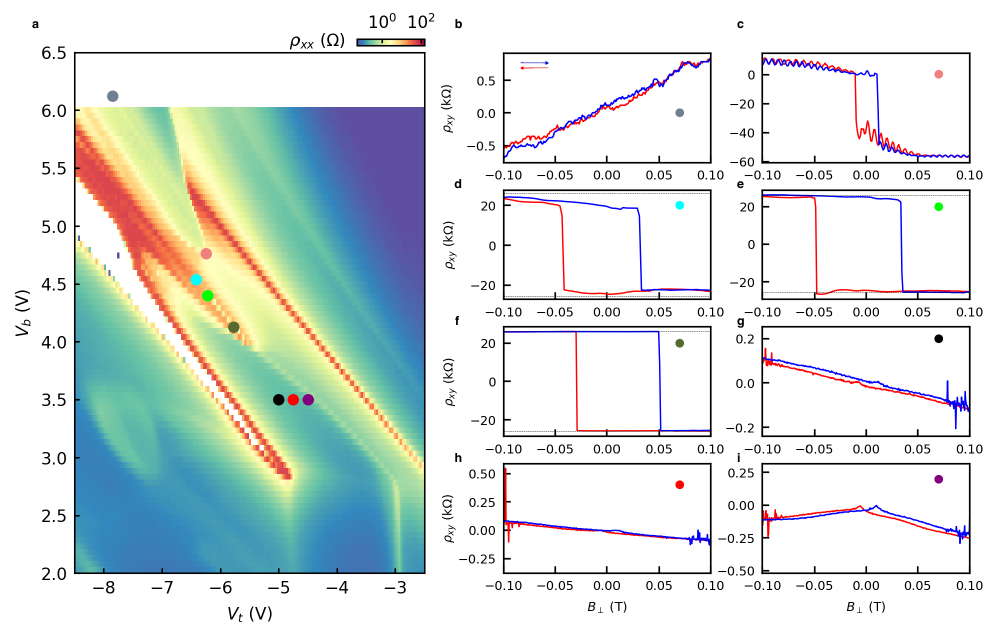}
\caption{\textbf{Assessment of magnetic hysteresis in the regime of an isolated conduction band.} \textbf{a}, map of $\rho_{xx}$ versus $V_t$ and $V_b$ with colored dots showing where each hysteresis loop is taken. Measurement of $\rho_{xy}$ vs $B_{\perp}$ swept in both directions (blue denotes the forward direction and red the backward), taken at \textbf{b}, ($V_t$,$V_b$)= ($-7.84$~V, $6.12$~V), marked by the grey dot in \textbf{(a)}, \textbf{c}, ($-6.25$~V, $4.76$~V), marked by the light red dot, \textbf{d}, ($-6.42$~V, $4.54$~V), marked by the cyan dot, \textbf{e}, ($-6.22$~V, $4.40$~V), marked by the lime dot, \textbf{f}, ($-5.77$~V, $4.12$~V), marked by the dark green dot, \textbf{g}, ($-5.00$~V, $3.50$~V), marked by the black dot, \textbf{h}, ($-4.75$~V, $3.50$~V), marked by the red dot, and \textbf{i}, ($-4.50$~V, $3.50$~V), marked by the purple dot. $B_{\parallel}=0$~T for all the measurements.
}
\label{fig:HysteresisMeasurements}
\end{figure*}

\begin{figure*}[t]
\includegraphics[width=0.95\textwidth]{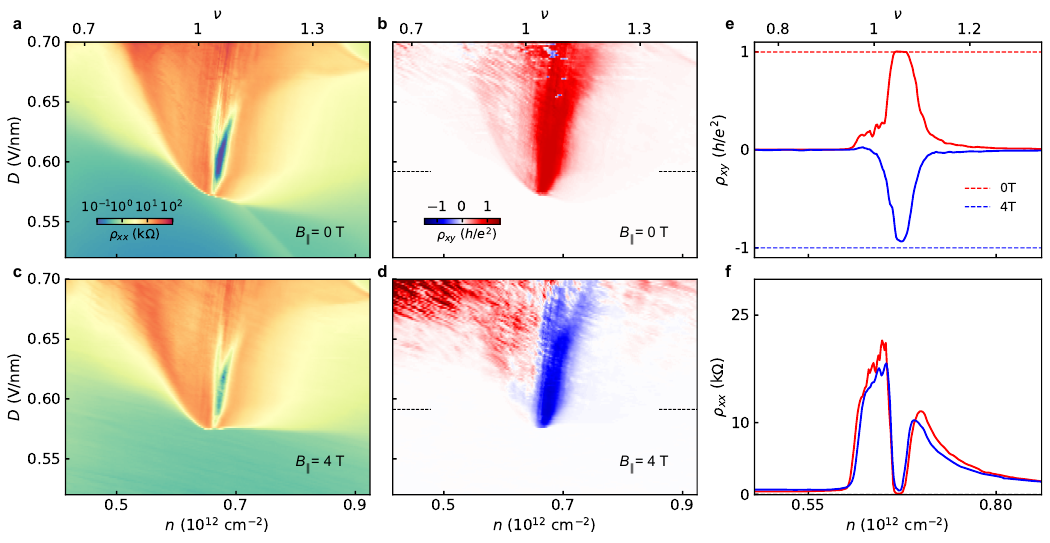}
\caption{\textbf{Integer quantum anomalous Hall state at different $B_\parallel$.} \textbf{a}, Map of $\rho_{xx}$ versus $D$ and $n$ taken at $B_\parallel=0$~T. \textbf{b}, Corresponding $\rho_{xy}$ to \textbf{(a)}. \textbf{c}, Map of $\rho_{xx}$ versus $D$ and $n$ taken at $B_\parallel=4$~T. \textbf{d}, Corresponding $\rho_{xy}$ to \textbf{(c)}. \textbf{e}, $\rho_{xy}$ linecuts along the black dashed lines in \textbf{(b)} and \textbf{(d)}, taken at $D=0.59$~V/nm. \textbf{f}, Corresponding $\rho_{xx}$ to \textbf{(e)}. $\rho_{xx}$ data shown in \textbf{(a-f)} is taken using $V_{xx2}$. All measurements are taken with $B_\perp=0$.
} 
\label{fig:IQAH0Tvs4T}
\end{figure*}

\FloatBarrier     

\newpage

\renewcommand{\figurename}{Supplementary Information Fig.}
\renewcommand{\thesubsection}{S\arabic{subsection}}
\setcounter{secnumdepth}{2}
\renewcommand{\thetable}{S\arabic{table}}
\setcounter{figure}{0} 
\setcounter{equation}{0}

\onecolumngrid
\newpage

\FloatBarrier           
\section*{Supplementary Information}
\FloatBarrier  

\subsection{Ginzburg-Landau analysis of ferromagnetic half metals with weak spin-orbit coupling}

In this Section we use a phenomenological toy model to investigate how spin-polarized but valley-unpolarized half metals respond to a weak (intrinsic) Kane-Mele spin-orbit coupling. We construct a Ginzburg-Landau free energy by following the hierarchy of energy scales present in rhombohedral graphene systems. 

Taking a spin-polarized metal as a starting point, the order parameters for our theory will be the spin-polarization vectors $\bm S_\pm$ in the two valleys ($\tau = \pm$) of graphene, which we assume uniform in space. Microscopically, we can think of $\bm S_\tau$ with components $S_\tau^j = \frac{1}{A} \sum_{\bm k, \sigma} \langle c^\dagger_{\tau \sigma} (\bm k) s^j c_{\tau \sigma}(\bm k) \rangle$, where $A$ is the system area, $j=x, y, z$ and $s^j$ is the corresponding spin Pauli matrix, $\sigma$ denotes the layer and sublattice degrees of freedom, and $c_{\tau \sigma}(\bm k) = \left( c_{\tau \sigma \uparrow}(\bm k), c_{\tau \sigma \downarrow}(\bm k) \right)^T$ collects the relevant electron annihilation operators. The direction of $\bm S_\pm$ describes the orientation of the spin polarization in each valley, while its magnitude corresponds to the associated spin polarization density. For a true ferromagnetic half metal, where all doped electrons have the same spin, the magnitudes $S_\pm = |\bm S_\pm| = n_e/2$, where $n_e$ is the electronic density. (The factor of 2 accounts for the two valleys, such that the total electronic density $n_e = |\bm S_+| + |\bm S_-|$. Similar reasoning holds for hole doping.) Note that when minority-flavor pockets are present (or when valence and conduction band carriers coexist) this exact mapping to the experimentally-controlled $n_e$ is lost.

\subsubsection{Isotropic theory}

We first consider the largest energy scales in the problem: the combined effect of band dispersion and long-range Coulomb repulsion (which neglects inter-valley exchange processes). These terms respect the SU(2)$_+ \times$ SU(2)$_-$ symmetry group, comprising independent spin rotations in each valley, and time-reversal symmetry which exchanges the two spin polarizations, $\bm S_+ \leftrightarrow - \bm S_-$. The most general free energy density with these constraints reads
\begin{equation}
\label{eq:free_energy_sym}
    F = \frac{\alpha}{2} \left( \bm S_+^2 + \bm S_-^2 \right) + \frac{\beta}{4} \left( \bm S_+^4 + \bm S_-^4 \right) + \frac{\gamma}{2} \bm S_+^2 \bm S_-^2 + \hdots ,
\end{equation}
where the ellipsis denote sixth- and higher-order terms. Spin polarization is nucleated when the quadratic coefficient $\alpha$ becomes negative. As we shall see the combined effect of $\beta$ and $\gamma$ is to bias the system either towards a valley-balanced solution (where both order parameters condense with the same magnitude) or towards valley imbalance (characterized by unequal $S_+$ and $S_-$).

Minimizing the free energy with respect to $S_+$ and $S_-$, we obtain a set of coupled cubic equations that admit two types of non-trivial solutions: valley balanced (VB), with $S_+ = S_- = \sqrt{\frac{-\alpha}{\beta + \gamma}}$, and valley imbalanced (VI), with $S_+ = \sqrt{\frac{-\alpha}{\beta}}$ and $S_-=0$ or vice-versa. These solutions have energy densities
\begin{align}
    F_{\rm VB} &= - \frac{\alpha^2}{2 (\beta + \gamma)} ~~~~ , ~~~~ F_{\rm VI} = - \frac{\alpha^2}{4 \beta}.
\end{align}
Valley balance thus prevails if $\gamma < \beta$, whereas valley imbalance occurs in the other limit. Here we assume that $\beta > 0$ but don't fix the sign of $\gamma$; stability of the theory to fourth-order requires $\gamma > -\beta$. 

We now add a Hund's coupling term, whose main role is to reduce the large SU(2)$_+ \times$ SU(2)$_-$ symmetry of the model to physical SU(2) spin rotations. A suite of experimental evidence~\cite{Zhou2021_2,Zhou2022_BBG} in rhombohedral graphene suggests that this term is ferromagnetic, $- J \bm S_+ \cdot \bm S_-$ with $J>0$, in order to promote ferromagnetism (i.e. aligned spin polarization in the two valleys) among the various configurations permitted by long-range Coulomb interactions. This phenomenological addition can be thought of as arising from inter-valley scattering due to lattice-scale electronic repulsion~\cite{Chatterjee2022intervalley}. 

Hund's coupling lowers the energy of the valley-balanced solution. Aligning the two vectors along an arbitrary axis and setting their magnitudes equal, $S_+ = S_- = S_0$, one obtains
\begin{equation}
    F_{\rm VB} = \left( \alpha - J \right) S_0^2 + \frac{\beta + \gamma}{2} S_0^4 .
    \label{eq:free_energy_VB_Hunds}
\end{equation}
Minimizing over $S_0$ yields the optimal value $S_0 = \sqrt{\frac{-\alpha + J}{\beta + \gamma}}$ and
\begin{align}
    F_{\rm VB} &= - \frac{(\alpha - J)^2}{2 (\beta + \gamma)}.
\end{align}
In contrast, Hund's coupling is inoperative in a fully valley-polarized phase due to one of the vectors $\bm S_\pm$ vanishing. We can however consider a partially valley-imbalanced solution, where a fraction of electrons are transferred from one valley to the other. Aligning the spins and taking $S_\pm = S_0 \pm \delta$ with a fixed total density $S_0$, we obtain
\begin{equation}
    F_{\rm VI} = \left( \alpha - J \right) S_0^2 + \frac{\beta + \gamma}{2} S_0^4 + \left(\alpha + J + (3 \beta  - \gamma) S_0^2  \right) \delta^2 + \mathcal{O}(\delta^4) ,
\label{eq:free_energy_VI_mixed}
\end{equation}
where we dropped terms that are fourth-order in $\delta$. Note that the first two terms in Eq.~\eqref{eq:free_energy_VI_mixed} are identical to Eq.~\eqref{eq:free_energy_VB_Hunds}. The system becomes unstable towards developing valley imbalance if the quadratic coefficient $\sim \delta^2$ becomes negative. Neglecting corrections to $S_0$ by a small $\delta$ near the onset of valley imbalance, the instability occurs when

\begin{equation}
   \frac{\gamma}{\beta} > 1 +  \frac{2 J}{|\alpha|}.
\end{equation}
As expected on symmetry grounds, Hund's coupling enlarges the regime of stability of the spin-polarized and valley-balanced solution, as compared to the SU(2)$_+ \times$SU(2)$_-$ theory (where imbalance onsets already for $\gamma/\beta > 1$).

\subsubsection{Spin-orbit coupling}

Graphene layers host weak Kane-Mele spin-orbit coupling~\cite{KaneMele2005Quantum}, which acts in the low-energy theory as $\sim \sigma^z \tau^z s^z$, where Pauli matrices $\sigma$, $\tau$ and $s$ respectively act on the sublattice, valley and spin degrees of freedom. In rhombohedral graphene near charge neutrality, the conduction/valence bands become strongly layer- and sublattice-polarized under the application of a perpendicular displacement field $D$---in which case the Kane-Mele SOC acts effectively like an Ising term $\sim \tau_z s_z$, with an opposite sign in the conduction and valence bands that also depends on the direction of $D$. 
Various experiments have estimated the scale of this intrinsic Kane-Mele term to a few tens of $\mu$eV~\cite{Arp2024,Sichau2019,Banszerus2020,Kurzmann2021}. While sub-leading compared to long-range Coulomb and Hund's interactions, an effective Ising SOC term of that scale can have significant effects on pairing as the BCS gap $\Delta \sim 1.76 k_B T_c \sim 15 ~ \mu$eV for $T_c \sim 100$ mK.

We thus consider the possible metallic ground states when Kane-Mele SOC is added to the theory:
\begin{equation}
\label{eq:free_energy}
    F = \frac{\alpha}{2} \left(\bm S_+^2 + \bm S_-^2 \right) + \frac{\beta}{4} \left(\bm S_+^4 + \bm S_-^4 \right) + \frac{\gamma}{2} \bm S_+^2 \bm S_-^2 - J \bm S_+ \cdot \bm S_- - \frac{\lambda}{2} (S_+^z - S_-^z) .
\end{equation}
The spin symmetry is now lowered and only comprises U(1)$_z$ rotations about the $z$ axis. There are two types of spin-polarized ground states: those where the spontaneous magnetization points in the graphene layers (easy-plane), thus spontaneously breaking the U(1)$_z$ symmetry, and those where the magnetization lies out of plane (easy-axis).

We first consider the easy-plane case and assume the magnetization develops in the $x$ direction without loss of generality. We parametrize the spin vectors with a polar angle $\theta$ measured from the $z$ axis: $\bm S_\pm = S_0 (\sin \theta, 0, \pm \cos \theta)$. In this ``spin-canted'' configuration, both valleys host the same in-plane magnetic moment, but opposite out-of-plane moments to satisfy Kane-Mele SOC~\cite{Koh2024trilayer}. Because the magnitudes $S_\pm$ are identical, this phase remains a half-metal (with two degenerate Fermi surfaces) for all $\theta$. Feeding this ansatz into Eq.~\eqref{eq:free_energy}, one gets
\begin{equation}
    \label{eq:free_energy_canted}
    F_{\rm canted} = \left(\alpha - J \right)  S_0^2 +  \frac{\beta + \gamma}{2} S_0^4 + 2 J S_0^2 \cos^2 \theta - \lambda S_0 \cos \theta .
\end{equation}
Minimizing over the canting angle $\theta$ yields two possible solutions. If Hund's coupling is strong enough ($4 J S_0 > \lambda$), one finds a non-trivial solution with $\cos \theta_0 = \frac{\lambda}{4 J S_0}$~\cite{Dong2025superconductivity}. Otherwise, the system remains spin-valley-locked at $\theta_0 = 0$. In bare graphene (i.e. without a proximitizing TMD susbtrate), the scale of $\lambda$ is small enough that spin-canting occurs at essentially all physically relevant densities~\cite{Arp2024, Patterson2025}. Plugging back $\theta_0$ in Eq.~\eqref{eq:free_energy_canted}, we get
\begin{equation}
    F_{\rm canted} = \left(\alpha - J \right)  S_0^2 +  \frac{\beta + \gamma}{2} S_0^4 - \frac{\lambda^2}{8 J}.
    \label{eq:optimal_canted}
\end{equation}
Spin canting thus enjoys a quadratic energy gain from Kane-Mele SOC, which is inversely proportional to the Hund's coupling that prefers parallel spin polarizations.

A bit more subtle is the alternative easy-axis option, where magnetism develops out-of-plane. In this case, Kane-Mele SOC acts oppositely to the spontaneously-generated magnetization in the two valleys. A valley imbalance is thus introduced, diagnosed in our model by a different length of the spin vectors. To describe this phase we take the ansatz $\bm S_\pm = (0, 0, S_0 \pm \delta)$, which introduces a valley imbalance $\delta$ while preserving the total polarization density. Plugging into the free energy Eq.~\eqref{eq:free_energy}, we have (up to second-order in $\delta$):
\begin{equation}
    F_{\rm VI} = \left( \alpha - J \right) S_0^2 + \frac{\beta + \gamma}{2} S_0^4 + \left(\alpha + J + (3 \beta  - \gamma) S_0^2  \right) \delta^2  - \lambda \delta .
\end{equation}
Contrary to the isotropic case, the system develops a valley imbalance due to the linear term enabled by Kane-Mele SOC (even if the quadratic coefficient is positive). Minimizing over $\delta$ yields 
$\delta_0 = \frac{\lambda}{2 \left( \alpha + J + (3 \beta - \gamma) S_0^2 \right)}$
and the optimal valley-imbalanced solution has energy density 
\begin{equation}
    F_{\rm VI} = \left( \alpha - J \right) S_0^2 + \frac{\beta + \gamma}{2} S_0^4 - \frac{\lambda^2}{4 \left( \alpha + J + (3 \beta - \gamma) S_0^2 \right)} .
    \label{eq:optimal_VI}
\end{equation}
We again find a quadratic energy gain due to Kane-Mele SOC. 

Comparing the spin-canted and valley-imbalanced solutions in Eqs.~\eqref{eq:optimal_canted} and \eqref{eq:optimal_VI}, valley imbalance is favored when
\begin{equation}
8 J > 4 \left( \alpha + J + (3 \beta - \gamma) S_0^2 \right) .
\end{equation}
Using the saddle-point value $S_0^2 = \frac{-\alpha + J}{\beta + \gamma}$ obtained from Eq.~\eqref{eq:free_energy_VB_Hunds}---again ignoring small corrections to it from $\delta$---the above inequality translates to
\begin{equation}
J (\gamma - \beta) > |\alpha| (\beta - \gamma).
\end{equation}
Again remember that $J, |\alpha|, \beta > 0$. There are two possible cases: if $\beta > \gamma$, the inequality can never be satisfied as the left-hand side is negative but the right-hand side positive. Spin canting thus wins in this limit. In the opposite limit $\gamma > \beta$, the inequality is always satisfied and valley imbalance prevails.

\subsubsection{Summary of the phase competition and numerics}

We have identified three regimes as a function of the competition between the fourth-order terms $\gamma$ and $\beta$:
\begin{itemize}
    \item (1) $\frac{\gamma}{\beta} < 1$: Valley balance always preferred (spin canting favored by Kane-Mele SOC)
    \item (2) $1 < \frac{\gamma}{\beta} < 1 + \frac{2J}{|\alpha|}$: Valley balance in the isotropic theory, but Kane-Mele SOC introduces a valley imbalance
    \item (3) $\frac{\gamma}{\beta} > 1 + \frac{2J}{|\alpha|}$: Valley imbalance present already in the isotropic theory $\rightarrow$ (generalized) quarter metal
\end{itemize}
Of most interest to experiment are the first two regimes, where a spin-polarized but valley-balanced half metal is selected by long-range Coulomb and Hund's coupling. Depending on the ratio of $\gamma$ and $\beta$, a weak Kane-Mele spin-orbit coupling term can induce a valley imbalance rather than lead to spin canting. This tendency increases when the system approaches the boundary between a half metal and a (generalized) quarter metal or three-quarter metal in its density and displacement-field tuned phase diagram---i.e., when approaching regime (3).

We now illustrate the phase competition by numerically minimizing the free energy density in Eq.~\eqref{eq:free_energy}. 
To perform numerics we recast the theory such that all parameters have energy units. We rescale $\bm S_\pm = \tilde{\bm S}_\pm/A_0$, where $\tilde{\bm S}_\pm$ describe a number of polarized electrons (without units) and $A_0$ is an area that we set for convenience to $A_0 = 1 \times 10^{-12}$ cm$^{2}$$ = 100$ nm$^2$, a natural scale for spin polarization in rhombohedral graphene. The Hund's coupling is estimated at $J \sim 200$ meV nm$^2$~\cite{Arp2024,Patterson2025}, which in these units becomes an energy scale $\sim 2$ meV. We consider a stronger coupling $|\alpha| \sim 1$ eV nm$^2$ for long-range Coulomb interactions, leading to an energy scale $\sim 10$ meV. To obtain a spin polarization of order $1$ in these units (or a spin polarization density of $\sim 10^{12}$ cm$^{-2}$) we take $\beta = 5$ meV and $\gamma$ of the same order, remembering that $S_0 = \sqrt{(|\alpha| + J)/(\beta + \gamma)}$.

\begin{figure*}[t]
\includegraphics[width=0.7\textwidth]{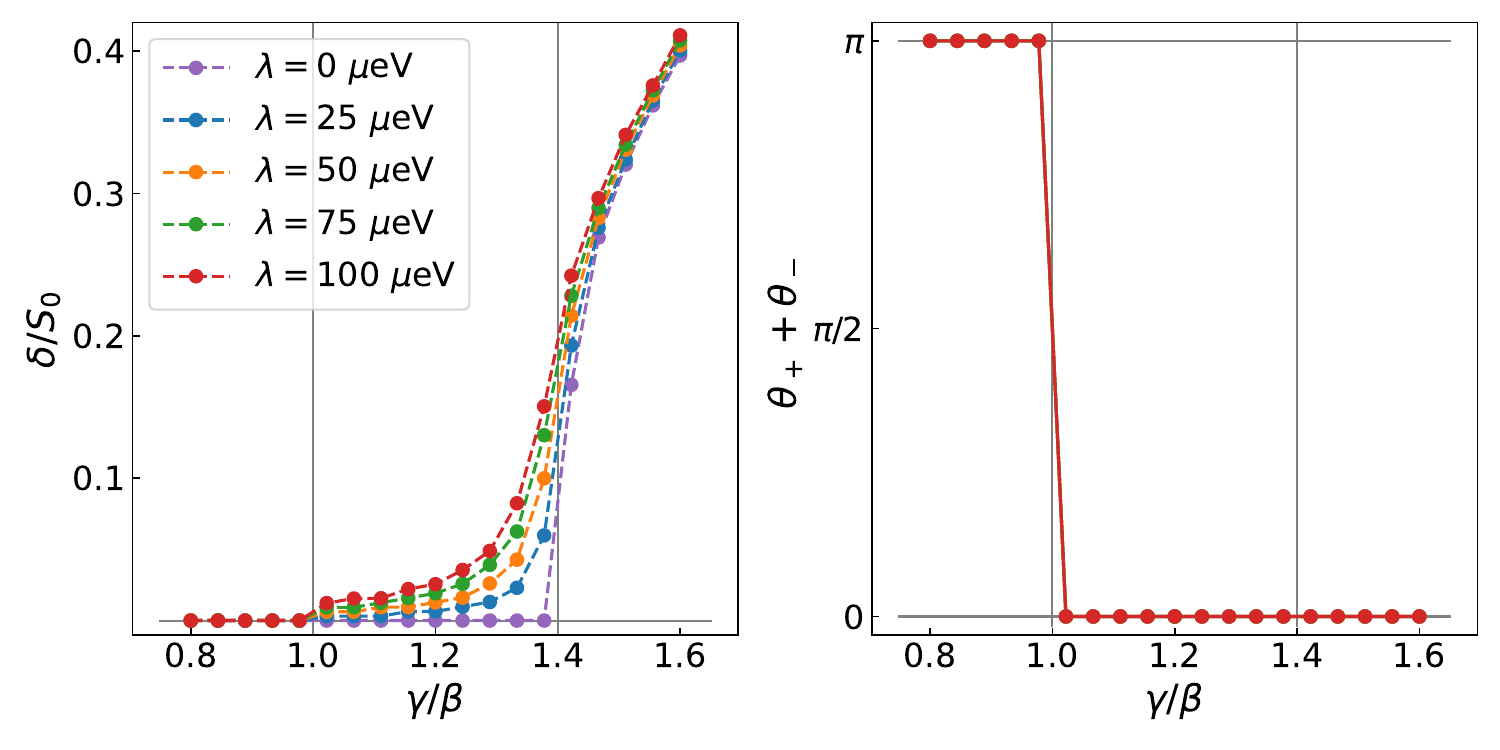} 
\caption{ \textbf{Spin canting vs valley imbalance at $B_\parallel = 0$}. Relative valley imbalance $\delta/S_0$ (left) and sum of the canting angles $\theta_\pm$ in the two valleys (right) as a function of the ratio $\gamma/\beta$ that captures tendency towards valley polarization. The spin-canted half metal is characterized by the condition $\theta_+ + \theta_- = \pi$. Results are obtained by numerically minimizing Eq.~\eqref{eq:free_energy}; numerical values of parameters used are mention in the text. The two gray vertical lines denote the onset of valley-imbalance in the cases with (at $\gamma/\beta = 1$) and without (at $\gamma/\beta = 1.4$) Kane-Mele SOC. Note that when $\lambda=0$ the canting angle (right panel) is ill-defined due to the SU(2) spin symmetry of the theory---we therefore did not include it.}
\label{fig:SM_theory_zerofield}
\end{figure*}

Results of the minimization are shown in Fig.~\ref{fig:SM_theory_zerofield} for different values of $\lambda$. The system indeed undergoes two transitions as a function of $\gamma/\beta$ as specified above: for $\gamma/\beta > 1$ Kane-Mele SOC favors valley imbalance over spin canting, while for $\gamma/\beta > 1 + \frac{2 J}{|\alpha|} = 1.4$ valley imbalance is preferred already at $\lambda = 0$.
The numerics confirm that the tendency towards developing a valley imbalance with $\lambda$ increases when the ratio $\gamma / \beta$ is tuned towards the threshold $ 1 + \frac{2J}{|\alpha|}$ where the spin-polarized half metal becomes a (partially) valley-imbalanced quarter metal or three quarter metal.

\subsubsection{In-plane magnetic fields}

We finally consider the fate of the two (easy-axis and easy-plane) solutions in the presence of an in-plane magnetic field $B_\parallel$ (aligned in the $-x$ direction). Neglecting orbital effects and focusing on the Zeeman coupling to the spin polarization, the free energy density of the system becomes (assuming a $g$-factor of $2$):
\begin{equation}
\label{eq:free_energy_Bfields}
    F = \frac{\alpha}{2} \left(\bm S_+^2 + \bm S_-^2 \right) + \frac{\beta}{4} \left( \bm S_+^4 + \bm S_-^4 \right) + \frac{\gamma}{2} \bm S_+^2 \bm S_-^2 - J \bm S_+ \cdot \bm S_- - \frac{\lambda}{2} (S_+^z - S_-^z) - \mu_B B_\parallel (S_+^x + S_-^x) .
\end{equation}
This theory is now too complicated for analytical solutions. General considerations lead to the following intuition: 

\begin{itemize}
    \item The spin-canted phase should initially be favored under $B_\parallel$ (with a linear energy gain) as it hosts an in-plane magnetic moment already at zero field. The canting angles will become closer to $\pi/2$ (pure in-plane alignment) when $\mu_B B_\parallel$ overwhelms $\lambda/2$, but the resulting phase remains valley balanced for all $B_\parallel$.
    
    \item The valley-imbalanced phase will acquire an in-plane moment due to $B_\parallel$, which rotates the polarization vectors towards the plane (generically by different polar angles). This process is depicted schematically in the inset of Fig.~\ref{fig:3}a of the main text. As the spins become more planar, the valley imbalance correspondingly diminishes. The initial energy gain should be quadratic at small $B_\parallel$ due to the lack of an in-plane moment at zero field.
\end{itemize}

\begin{figure*}[t]
\includegraphics[width=\textwidth]{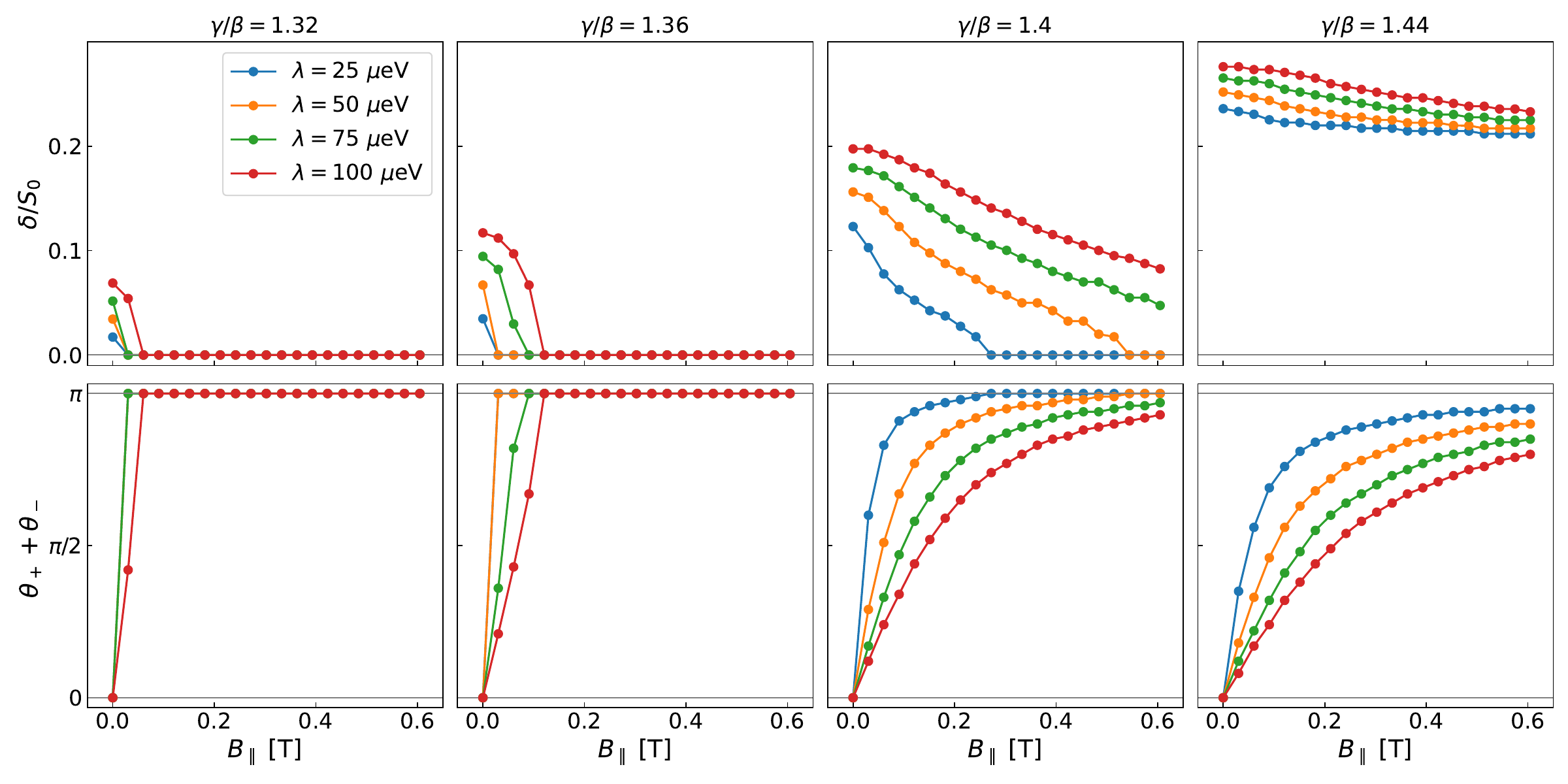} 
\caption{ \textbf{Spin canting vs valley imbalance as a function of in-plane magnetic field $B_\parallel$}. Relative valley imbalance $\delta/S_0$ (top panels) and sum of the canting angles $\theta_\pm$ in the two valleys (bottom panels) as a function of in-plane field $B_\parallel$. Results are obtained by numerically minimizing Eq.~\eqref{eq:free_energy_Bfields}. From left to right, we increase the ratio of fourth-order terms $\gamma/\beta$ that dictates the tendency towards valley imbalance. Here $J/|\alpha| = 0.2$ such that the critical value separating regimes (2) and (3) is $(\gamma/\beta)_{\rm critical} = 1.4$. For $\gamma/\beta$ smaller than the critical value, valley imbalance is favored at low fields but quickly loses out in favor of spin canting at a critical field $B_\parallel^c$. In contrast, beyond $(\gamma/\beta)_{\rm critical}$ the system remains in a valley-imbalanced metal for all $B_\parallel$. Precisely at the critical point, Coulomb and Hund's interactions do not have a preference for either phase---the critical field $B_\parallel^c$ where canting is restored is thus only set by $\lambda/2$, the energy scale of the easy-axis spin anisotropy.}
\label{fig:SM_theory}
\end{figure*}

To bolster this intuition we turn to a numerical minimization of Eq.~\eqref{eq:free_energy_Bfields} for generic spin-polarization vectors in the $xz$ plane, $\bm S_\pm = S_\pm (\sin \theta_\pm , 0, \cos \theta_\pm)$, as shown in Fig.~\ref{fig:SM_theory}. We use the same numerical values for $J$, $\alpha$, $\beta$ as in the previous section, and consider $\gamma/\beta > 1$ to investigate the fate of valley imbalance as a function of $B_\parallel$.

As $B_\parallel$ is increased from $0$, the (out-of-plane) spins start to tilt towards the graphene plane, and the valley polarization (top panels) correspondingly goes down. There are two distinct scenarios for how that process unfolds. In regime (2), a continuous phase transition occurs at a critical field $B_\parallel^c$, where the valley imbalance vanishes and the polar angles lock to $\theta_+ + \theta_- = \pi$ (bottom panels). In other words, sufficiently strong $B_\parallel$ restores spin canting. Note that the critical field $B_\parallel^c$ is generally much smaller than the spin anisotropy scale set by $\lambda/2$. This behavior arises because in regime (2) Coulomb and Hund's interactions conspire to favor a half-metal state, thus producing an energy penalty associated with valley imbalance, which the spin-canted phase does not suffer from. At the transition point (here for $\gamma/\beta=1.4$), Coulomb and Hund's terms compete to a draw and neither a half-metal nor a quarter-metal state is preferred. In this case, the competition is entirely set by the Zeeman and SOC terms, reflected by the scaling $B_\parallel^c \approx \lambda/2$. In contrast, in regime (3) where the system prefers to be in a (generalized) quarter-metal state, the spins rotate towards the plane but spin canting is never restored. Instead, the spin polarizations in the two valleys remain different, but progressively align with the direction of $B_\parallel$.

\subsection{Additional measurements characterizing superconductivity}

Below are the measurements used to extract the critical temperature of superconductivity, and an analysis of the nonzero saturation resistance.

\begin{figure*}[h]
\includegraphics[width=0.95\textwidth]{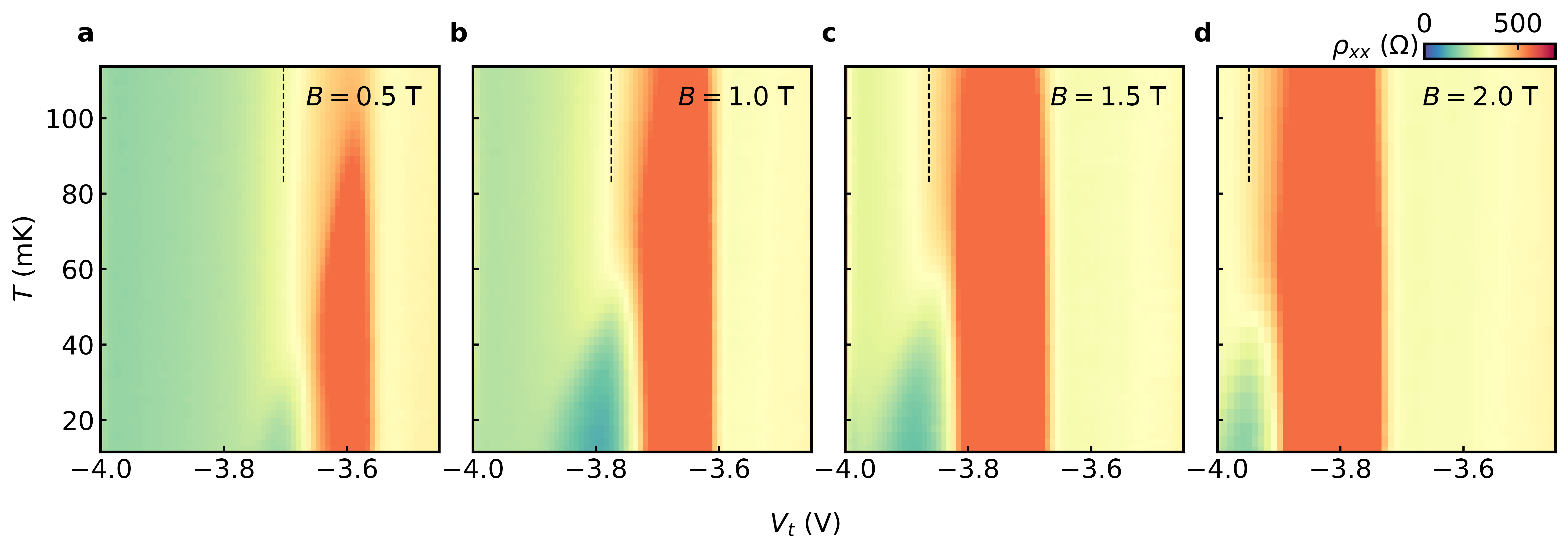} 
\caption{\textbf{Temperature dependence at $V_b$ = 3.1 V}.  $\rho_{xx}$ versus $V_t$ and $T$ taken at \textbf{a}, $B_{\parallel}=0.5$~T, \textbf{b}, $B_\parallel$ = 1.0 T, \textbf{c},  $B_\parallel$ = 1.5 T, \textbf{d}, $B_\parallel$ = 2.0 T. The dashed lines denote where the line cuts were taken in Extended Data Fig.~\ref{fig:TC_Vb_B}b.
}
\label{fig:TvsVt_3p1V}
\end{figure*}

\begin{figure*}[t]
\includegraphics[width=0.95\textwidth]{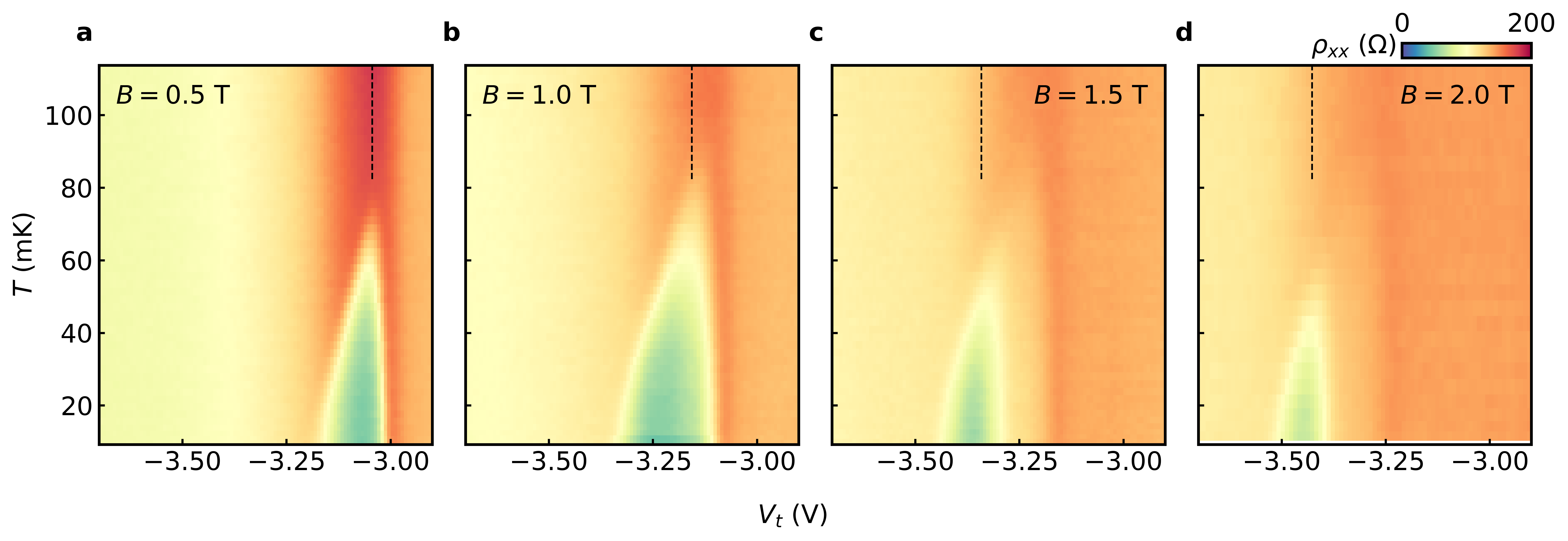} 
\caption{\textbf{Temperature dependence at $V_b$ = 2.1 V}.  $\rho_{xx}$ versus $V_t$ and $T$ taken at \textbf{a}, $B_{\parallel}=0.5$~T,  \textbf{b}, $B_\parallel$ = 1.0 T,  \textbf{c},  $B_\parallel$ = 1.5 T,  \textbf{d}, $B_\parallel$ = 2.0 T. The dashed lines denote where the line cuts were taken in Extended Data Fig.~\ref{fig:TC_Vb_B}c.
}
\label{fig:TvsVt_2p1V}
\end{figure*}

\begin{figure*}[t]
\includegraphics[width=0.95\textwidth]{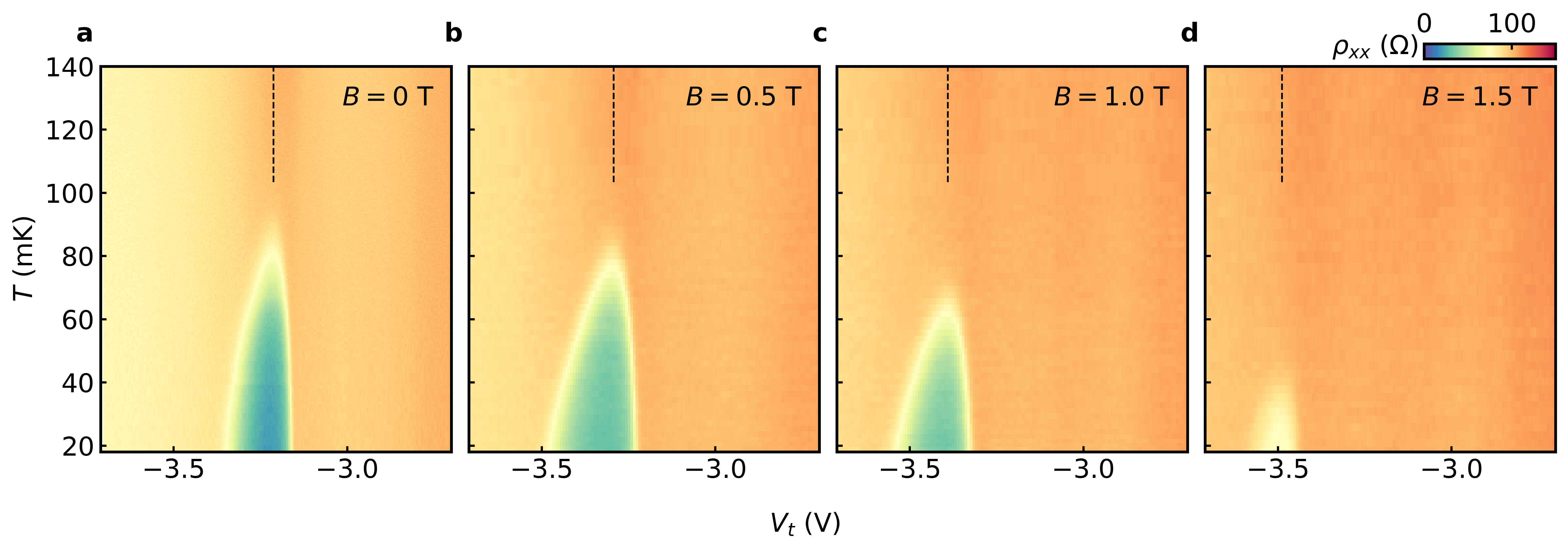} 
\caption{\textbf{Temperature dependence at $V_b$ = 1.4 V}.  $\rho_{xx}$ versus $V_t$ and $T$ taken at \textbf{a}, $B_{\parallel}=0$~T,  \textbf{b}, $B_\parallel$ = 0.5 T, \textbf{c},  $B_\parallel$ = 1.0 T, \textbf{d}, $B_\parallel$ = 1.5 T. The dashed lines denote where the line cuts were taken in Fig.~\ref{fig:3}b.  
}
\label{fig:TvsVt_1p4V}
\end{figure*}

\begin{figure*}[t]
\includegraphics[width=0.95\textwidth]{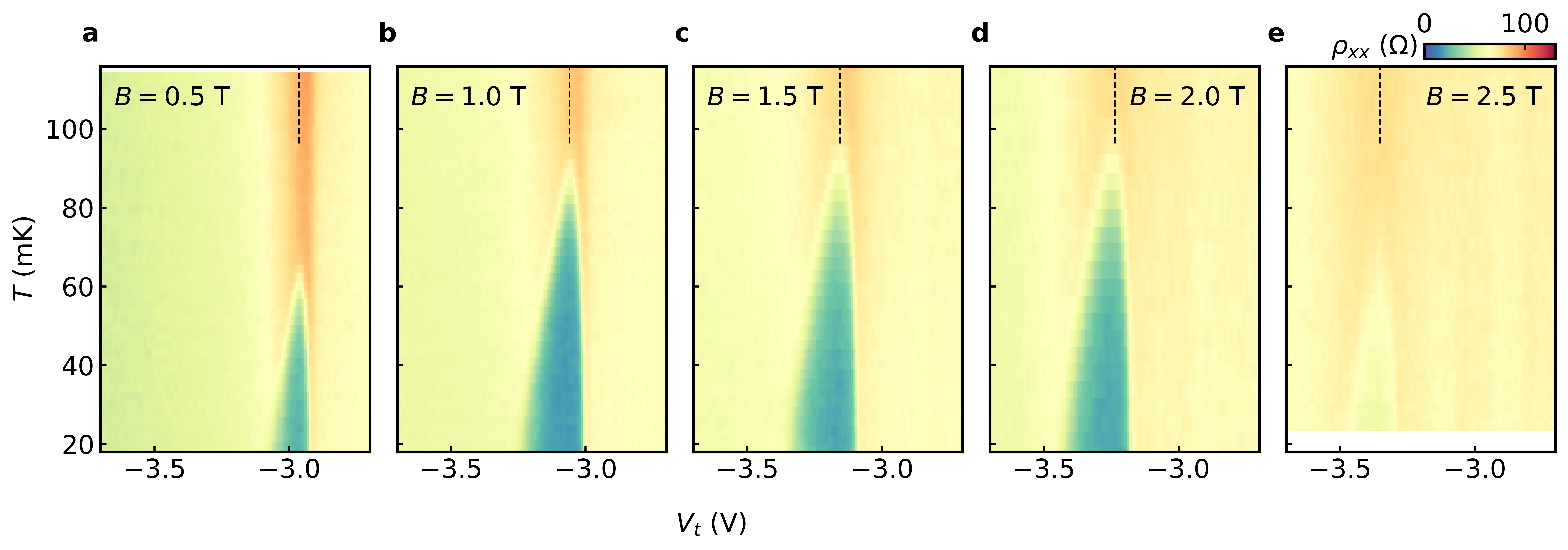} 
\caption{\textbf{Temperature dependence at $V_b$ = 0 V}.  $\rho_{xx}$ versus $V_t$ and $T$ taken at \textbf{a}, $B_{\parallel}=0.5$~T, \textbf{b}, $B_\parallel$ = 1.0 T, \textbf{c},  $B_\parallel$ = 1.5 T, \textbf{d}, $B_\parallel$ = 2.0 T, \textbf{e}, $B_\parallel$ = 2.5 T, The dashed lines denote where the line cuts were taken in Fig.~\ref{fig:3}c. 
}
\label{fig:TvsVt_0V}
\end{figure*}

\begin{figure*}[t]
\includegraphics[width=0.95\textwidth]{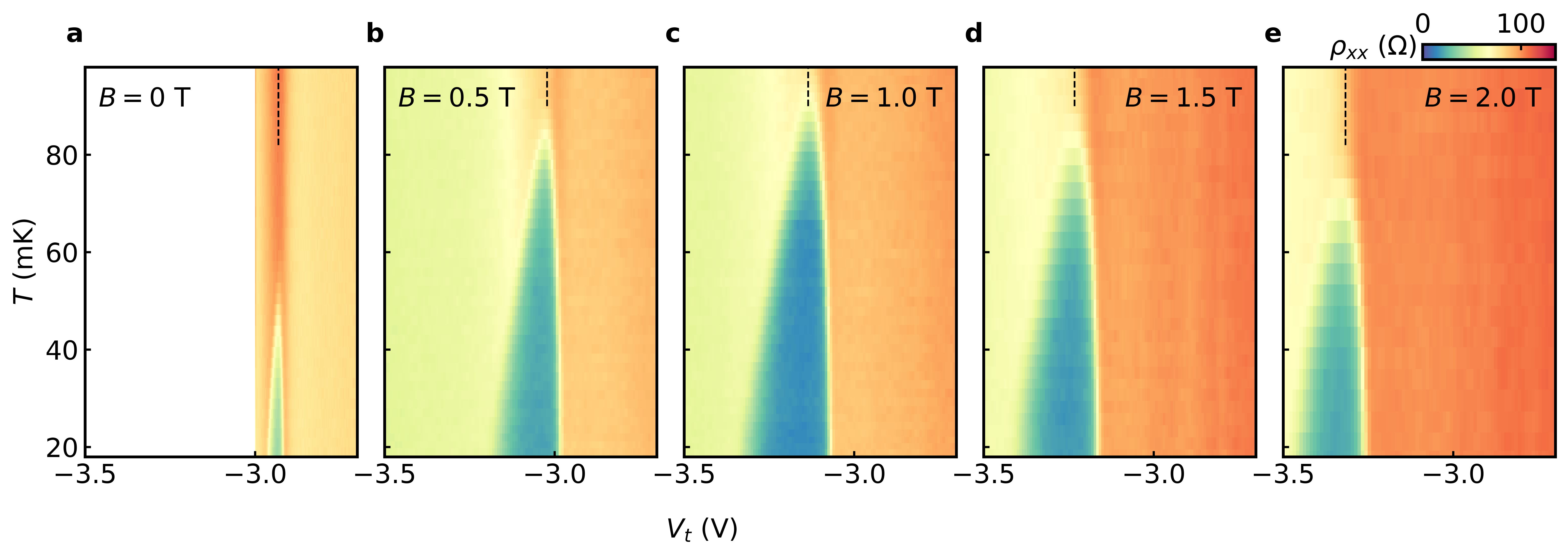} 
\caption{\textbf{Temperature dependence at $V_b$ = -0.95 V}.  $\rho_{xx}$ versus $V_t$ and $T$ taken at  \textbf{a}, $B_{\parallel}=0$~T,  \textbf{b}, $B_\parallel$ = 0.5 T, \textbf{c}, $B_\parallel$ = 1.0 T, \textbf{d}, $B_\parallel$ = 1.5 T, \textbf{e}, $B_\parallel$ = 2.0 T. The dashed lines denote where the line cuts were taken in Fig.~\ref{fig:3}d. 
}
\label{fig:TvsVt_n0p95V}
\end{figure*}

\begin{figure*}[t]
\includegraphics[width=0.95\textwidth]{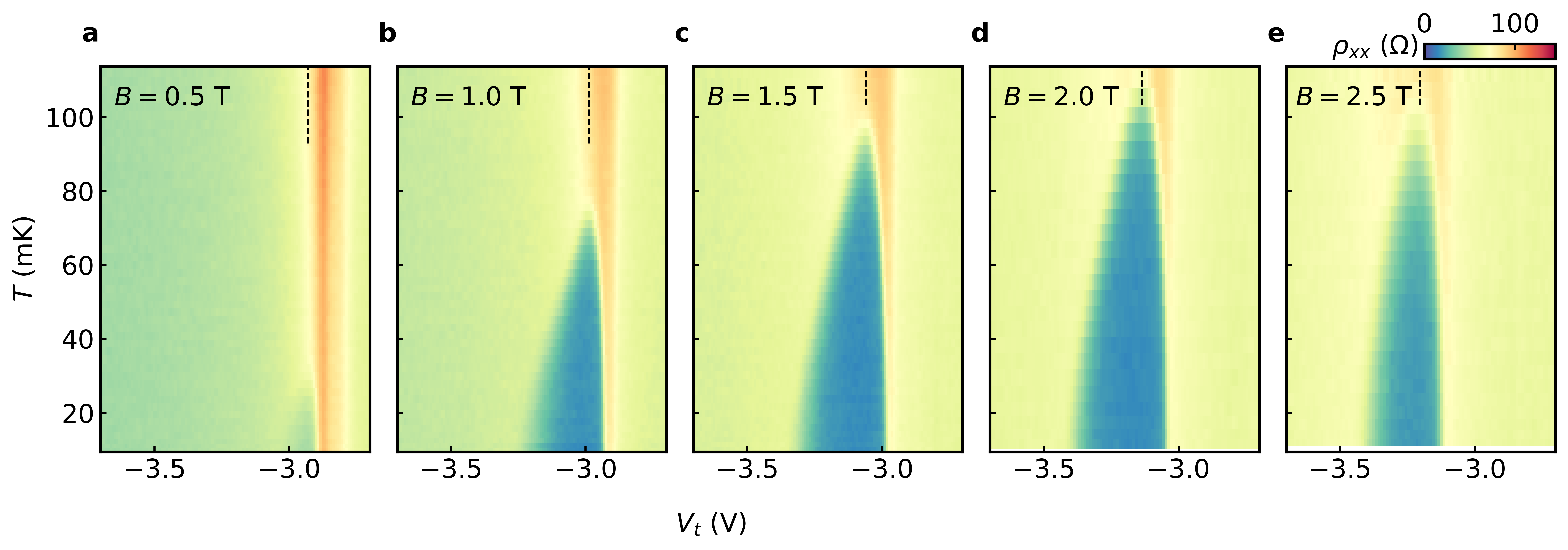} 
\caption{\textbf{Temperature dependence at $V_b$ = -3.0 V}. $\rho_{xx}$ versus $V_t$ and $T$ taken at \textbf{a},  $B_{\parallel}=0.5$~T,  \textbf{b}, $B_\parallel$ = 1.0 T, \textbf{c}, $B_\parallel$ = 1.5 T, \textbf{d}, $B_\parallel$ = 2.0 T, \textbf{e}, $B_\parallel$ = 2.5 T. The dashed lines denote where the line cuts were taken in Fig.~\ref{fig:3}e. 
}
\label{fig:TvsVt_n3V}
\end{figure*}

\begin{figure*}[t]
\includegraphics[width=0.95\textwidth]{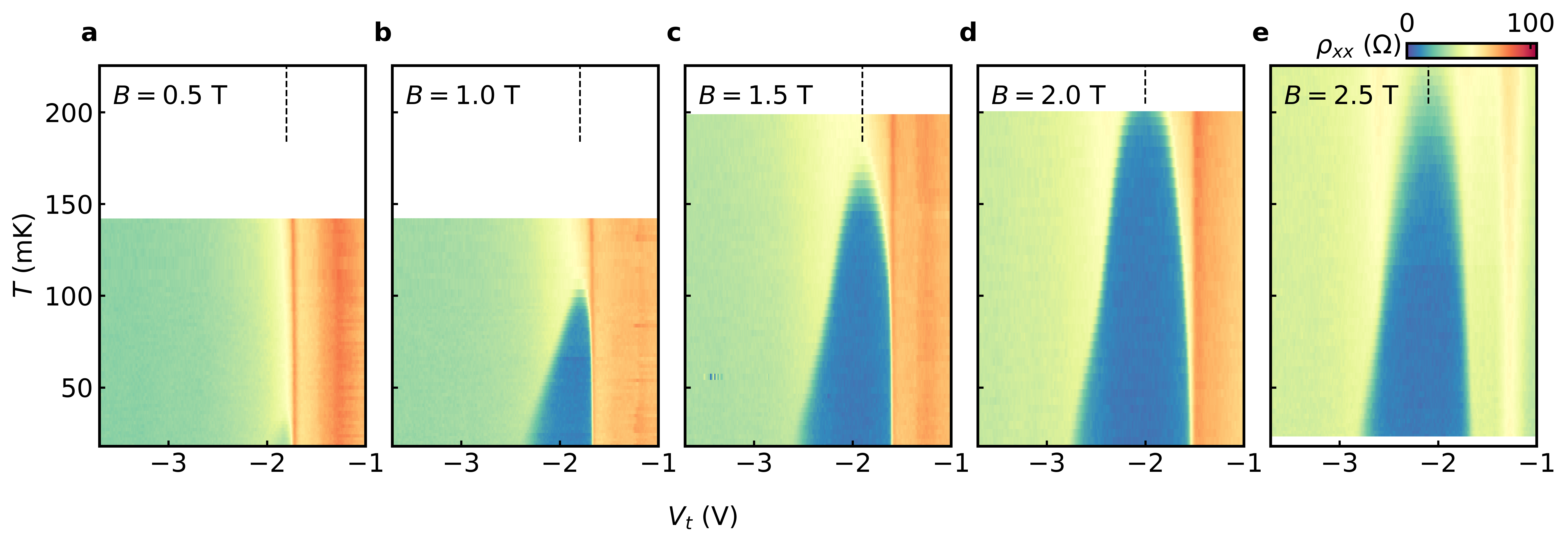} 
\caption{\textbf{Temperature dependence at $V_b$ = -4.75 V}. $\rho_{xx}$ versus $V_t$ and $T$ taken at  \textbf{a}, $B_{\parallel}=0.5$~T,  \textbf{b}, $B_\parallel$ = 1.0 T , {c}, $B_\parallel$ = 1.5 T,  \textbf{d}, $B_\parallel$ = 2.0 T, \textbf{e}, $B_\parallel$ = 2.5 T . The dashed lines denote where the line cuts were taken in Extended Data Fig.~\ref{fig:TC_Vb_B}d.
}
\label{fig:TvsVt_n4p75V}
\end{figure*}

\begin{figure*}[t]
\includegraphics[width=0.95\textwidth]{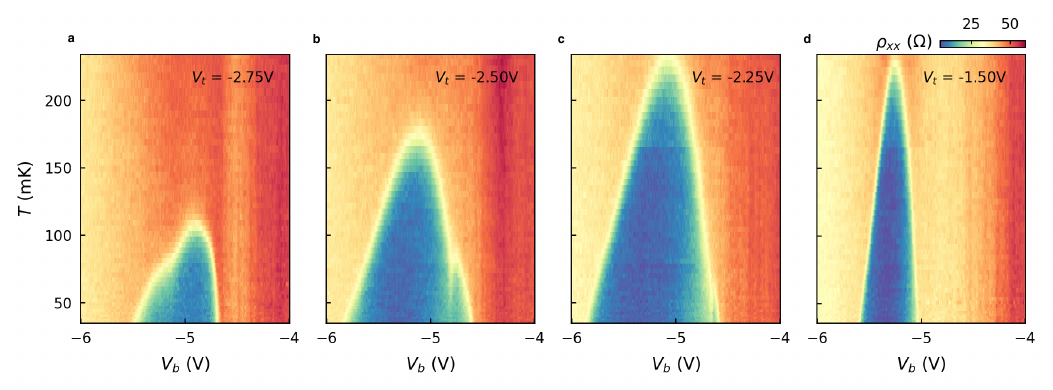} 
\caption{\textbf{Temperature dependence at $B_{\parallel}=4$~T}. $\rho_{xx}$ versus $V_b$ and $T$ with $B_{\parallel}=4$~T and taken at \textbf{a}, $V_t=-2.75$~V, \textbf{b}, $V_t=-2.5$~V, \textbf{c}, $V_t=-2.25$~V, and  \textbf{d}, $V_t=-1.5$~V.
}
\label{fig:TvsVt_4Tesla}
\end{figure*}

\begin{figure*}[t]
\includegraphics[width=0.95\textwidth]{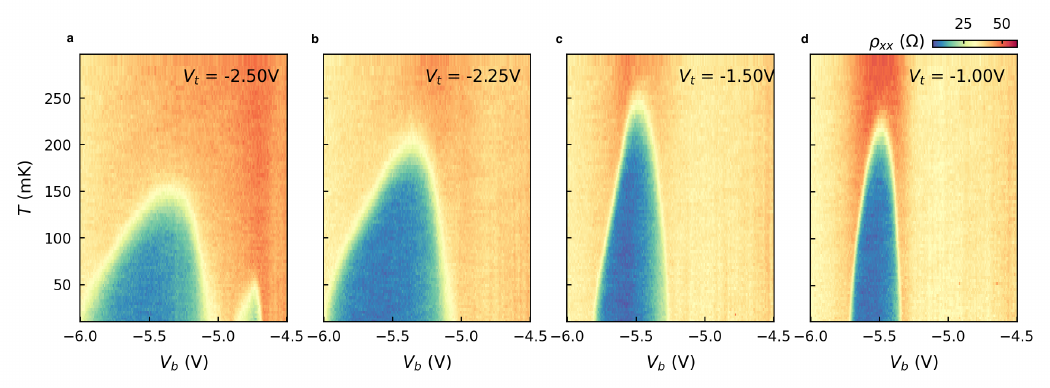} 
\caption{\textbf{Temperature dependence at $B_{\parallel}=6T$.} $\rho_{xx}$ versus $V_b$ and $T$ with $B_{\parallel}=6$~T  taken at \textbf{a}, $V_t=-2.5$~V, \textbf{b}, $V_t=-2.25$~V, \textbf{c}, $V_t=-1.5$~V, and  \textbf{d}, $V_t=-1$~V.
}
\label{fig:TvsVt_6Tesla}
\end{figure*}

\begin{figure*}[t]
\includegraphics[width=0.95\textwidth]{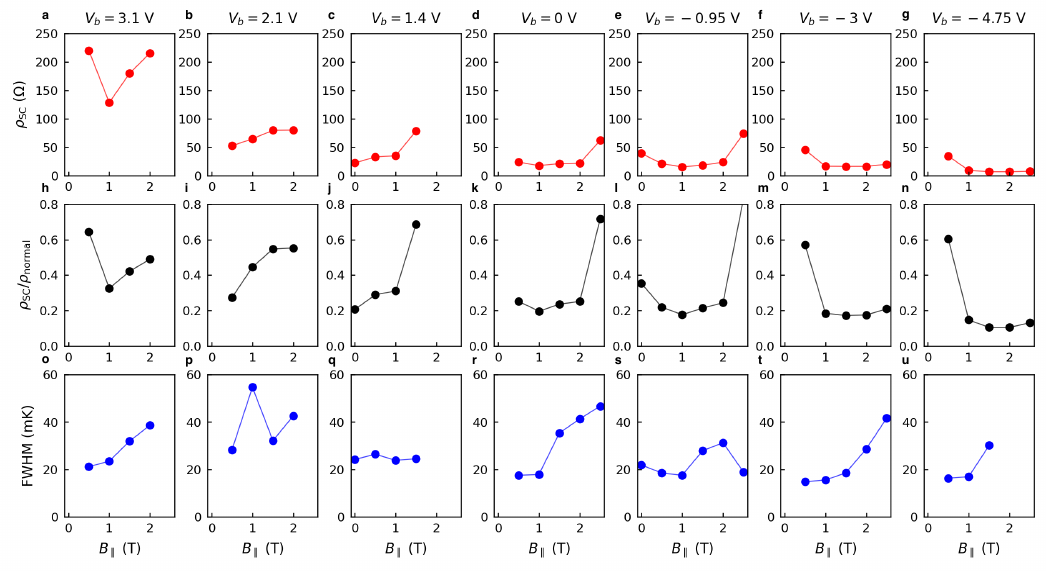} 
\caption{\textbf{Evolution of superconducting transition parameters with $B_{\parallel}$}. $\rho_{SC}$ (minimum resistivity in the superconducting state) versus $B_{\parallel}$ taken at \textbf{a}, $V_b=3.10$~V, \textbf{b}, $V_b=2.10$~V, \textbf{c}, $V_b=1.40$~V, \textbf{d}, $V_b=0.00$~V, \textbf{e}, $V_b=-0.95$~V, \textbf{f}, $V_b=-3.00$~V, and  \textbf{g}, $V_b=-4.75$~V. $\rho_{SC}$/$\rho_{normal}$ ($\rho_{normal}$ is the normal state resistivity) versus $B_{\parallel}$ taken at \textbf{h}, $V_b=3.10$~V,  \textbf{i}, $V_b=2.10$~V, \textbf{j}, $V_b=1.40$~V, \textbf{k}, $V_b=0.00$~V, \textbf{l}, $V_b=-0.95$~V, \textbf{m}, $V_b=-3.00$~V, and
\textbf{n}, $V_b=-4.75$~V. Full width half maximum (FWHM) of the rolling standard deviation around the superconducting transition versus $B_{\parallel}$ taken at \textbf{o}, $V_b=3.10$~V, \textbf{p}, $V_b=2.10$~V, \textbf{q}, $V_b=1.40$~V, \textbf{r}, $V_b=0.00$~V, \textbf{s}, $V_b=-0.95$~V, \textbf{t}, $V_b=-3.00$~V and \textbf{u}, $V_b=-4.75$~V. 
}
\label{fig:RscRnorm}
\end{figure*}

\end{document}